\documentstyle[12pt]{article}

\renewcommand{\theequation}{\thesection.\arabic{equation}}

\newcommand \beq{\begin{eqnarray}}
\newcommand \eeq{\end{eqnarray}}
\begin{document}
\input epsf


\def\nbfepsilon{\mbox{\boldmath$\epsilon$}}
\def\bfgrad{\mbox{\boldmath$\grad$}}
\def\bfgamma{\mbox{\boldmath$\gamma$}}
\def\bfcalA{\mbox{\boldmath${\cal A}$}}
\def\bfcalS{\mbox{\boldmath${\cal S}$}}
\def\bfp{\mbox{\boldmath$p$}}
\def\bfv{\mbox{\boldmath$v$}}
\def\bfj{\mbox{\boldmath$j$}}
\def\bfhp{\mbox{\boldmath$\hat p$}}
\def\bfei{\mbox{\boldmath$e_i$}}
\def\bfe{\mbox{\boldmath$e$}}
\def\bfej{\mbox{\boldmath$e_j$}}
\def\bfk{\mbox{\boldmath$k$}}
\def\bfq{\mbox{\boldmath$q$}}
\def\bfR{\mbox{\boldmath$R$}}
\def\bfC{\mbox{\boldmath$C$}}
\def\bfR{\mbox{\boldmath$R$}}
\def\bfX{\mbox{\boldmath$X$}}
\def\bfx{\mbox{\boldmath$x$}}
\def\bfE{\mbox{\boldmath$E$}}
\def\bfB{\mbox{\boldmath$B$}}
\def\bfy{\mbox{\boldmath$y$}}
\def\bfr{\mbox{\boldmath$r$}}
\def\rmRe{\mbox{\rm$Re$}}
\def\rmIm{\mbox{\rm$ImR$}}
\def\vp{\mbox{$\bf v\cdot p$}}
\def\vq{\mbox{$\bf v\cdot q$}}
\def\vpq{\mbox{$\bf v\cdot(p+ q)$}}

\def\bfgamma{\mbox{\boldmath$\gamma$}}
\def\bfalpha{\mbox{\boldmath$\alpha$}}
\def\bfsigma{\mbox{\boldmath$\sigma$}}
\def\bfalpha{\mbox{\boldmath$\alpha$}}
\def\bfsigma{\mbox{\boldmath$\sigma$}}
\def\bfSigma{\mbox{\boldmath$\Sigma$}}
\def\bfepsilon{\mbox{\boldmath$\epsilon$}}

\def\square{\hbox{{$\sqcup$}\llap{$\sqcap$}}}   
\def\grad{\nabla}                               
\def\del{\partial}                              

\def\frac#1#2{{#1 \over #2}}
\def\smallfrac#1#2{{\scriptstyle {#1 \over #2}}}
\def\half{\ifinner {\scriptstyle {1 \over 2}}
   \else {1 \over 2} \fi}

\def\bra#1{\langle#1\vert}              
\def\ket#1{\vert#1\rangle}              

\def\simge{\mathrel{%
   \rlap{\raise 0.511ex \hbox{$>$}}{\lower 0.511ex \hbox{$\sim$}}}}
\def\simle{\mathrel{
   \rlap{\raise 0.511ex \hbox{$<$}}{\lower 0.511ex \hbox{$\sim$}}}}


\def\parenbar#1{{\null\!                        
   \mathop#1\limits^{\hbox{\fiverm (--)}}       
   \!\null}}                                    
\def\nunubar{\parenbar{\nu}}
\def\ppbar{\parenbar{p}}


\def\buildchar#1#2#3{{\null\!                   
   \mathop#1\limits^{#2}_{#3}                   
   \!\null}}                                    
\def\overcirc#1{\buildchar{#1}{\circ}{}}


\def\slashchar#1{\setbox0=\hbox{$#1$}           
   \dimen0=\wd0                                 
   \setbox1=\hbox{/} \dimen1=\wd1               
   \ifdim\dimen0>\dimen1                        
      \rlap{\hbox to \dimen0{\hfil/\hfil}}      
      #1                                        
   \else                                        
      \rlap{\hbox to \dimen1{\hfil$#1$\hfil}}   
      /                                         
   \fi}                                         %


\def\subrightarrow#1{
  \setbox0=\hbox{
    $\displaystyle\mathop{}
    \limits_{#1}$}
  \dimen0=\wd0
  \advance \dimen0 by .5em
  \mathrel{
    \mathop{\hbox to \dimen0{\rightarrowfill}}
       \limits_{#1}}}                           

\def\real{\mathop{\rm Re}\nolimits}     
\def\imag{\mathop{\rm Im}\nolimits}     

\def\tr{\mathop{\rm tr}\nolimits}       
\def\Tr{\mathop{\rm Tr}\nolimits}       
\def\Det{\mathop{\rm Det}\nolimits}     

\def\mod{\mathop{\rm mod}\nolimits}     
\def\wrt{\mathop{\rm wrt}\nolimits}     

%
\def\journal#1#2#3#4{\ {#1}{\bf #2} ({#3})\  {#4}}

\def\AdvPhys{\journal{Adv.\ Phys.}}
\def\AnnPhys{\journal{Ann.\ Phys.}}
\def\EurophysLett{\journal{Europhys.\ Lett.}}
\def\JApplPhys{\journal{J.\ Appl.\ Phys.}}
\def\JMathPhys{\journal{J.\ Math.\ Phys.}}
\def\LettNuovoCimento{\journal{Lett.\ Nuovo Cimento}}
\def\Nature{\journal{Nature}}
\def\NPA{\journal{Nucl.\ Phys.\ {\bf A}}}
\def\NPB{\journal{Nucl.\ Phys.\ {\bf B}}}
\def\NuovoCimento{\journal{Nuovo Cimento}}
\def\Physica{\journal{Physica}}
\def\PLA{\journal{Phys.\ Lett.\ {\bf A}}}
\def\PLB{\journal{Phys.\ Lett.\ {\bf B}}}
\def\PR{\journal{Phys.\ Rev.}}
\def\PRC{\journal{Phys.\ Rev.\ {\bf C}}}
\def\PRD{\journal{Phys.\ Rev.\ {\bf D}}}
\def\PRB{\journal{Phys.\ Rev.\ {\bf B}}}
\def\PRL{\journal{Phys.\ Rev.\ Lett.}}
\def\PhysRept{\journal{Phys.\ Repts.}}
\def\ProcNatlAcadSci{\journal{Proc.\ Natl.\ Acad.\ Sci.}}
\def\ProcRoySoc{\journal{Proc.\ Roy.\ Soc.\ London Ser.\ A}}
\def\RevModPhys{\journal{Rev.\ Mod.\ Phys. }}
\def\Science{\journal{Science}}
\def\SovPhysJETP{\journal{Sov.\ Phys.\ JETP }}
\def\SovPhysJETPLett{\journal{Sov.\ Phys.\ JETP Lett. }}
\def\SovJNuclPhys{\journal{Sov.\ J.\ Nucl.\ Phys. }}
\def\SovPhysDoklady{\journal{Sov.\ Phys.\ Doklady}}
\def\ZPhys{\journal{Z.\ Phys. }}
\def\ZPhysA{\journal{Z.\ Phys.\ A}}
\def\ZPhysB{\journal{Z.\ Phys.\ B}}
\def\ZPhysC{\journal{Z.\ Phys.\ C}}


\begin{titlepage}
\begin{flushright} {Saclay-T97/052} 
\end{flushright}

\vspace*{0.2cm}
\begin{center}
\baselineskip=18pt {\Large THE BLOCH-NORDSIECK PROPAGATOR}

{\Large  AT FINITE TEMPERATURE}

 \vskip0.5cm Jean-Paul
BLAIZOT\footnote{CNRS}
  and Edmond IANCU\footnote{CNRS}
\\ {\it Service de Physique 
Th\'eorique\footnote{Laboratoire de la Direction des Sciences de
la Mati\`ere du Commissariat \`a l'Energie Atomique},
CE-Saclay \\ 91191 Gif-sur-Yvette, France}


\end{center}

\vskip 1cm
\vskip 1cm
\begin{abstract}

We have shown recently that the resummation of soft photon
contributions leads to  a $non$-$exponential$  decay of the
fermion excitations in hot QED plasmas.
The retarded propagator of a massless fermion was found to
behave as  $S_R(t\gg 1/gT)\sim
\exp\{-\alpha T t \,[\,\ln\omega_pt + { C}]\}$,
where $\omega_p=gT/3$ is the plasma frequency, $\alpha=g^2/4\pi$,
and $C$ is a  constant, independent of $g$, which was left undefined.
This term is computed in this paper. In gauges with unphysical 
degrees of freedom, it is gauge-fixing
independent  provided an infrared regulator is introduced 
in the gauge sector.  We also extend our analysis
to hot QCD and express the quark and gluon propagators in the 
form of three-dimensional Euclidean functional integrals which may be
evaluated on the lattice.

\end{abstract}

\vskip 2.6cm


\end{titlepage}


\baselineskip=18pt

\section{Introduction}
\setcounter{equation}{0}

\begin{figure}
\protect \epsfxsize=16.cm{\centerline{\epsfbox{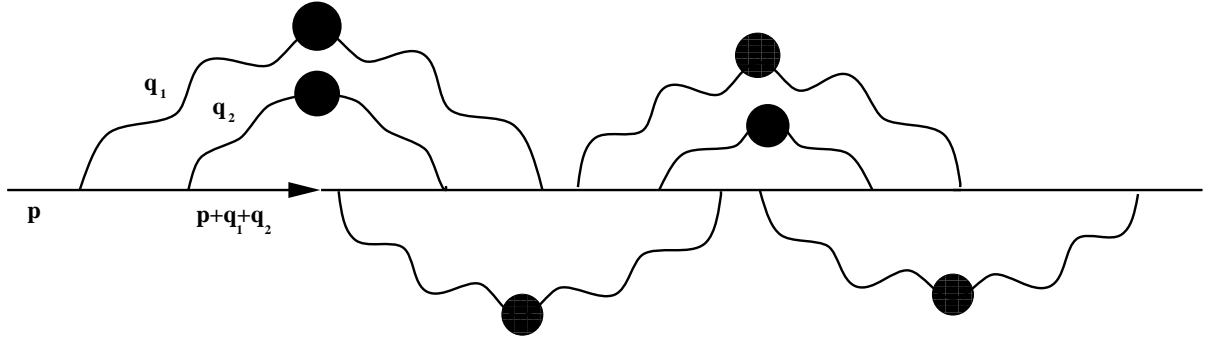}}}
	 \caption{A generic $n$-loop diagram (here, $n=6$)
which is responsible for infrared divergences in perturbation theory.
All the photon lines are soft and dressed by the hard thermal loop.
The fermion line is hard and nearly on-shell.}
\label{Nloop}
\end{figure}

The Bloch-Nordsieck (BN) approximation \cite{BN37} offers an economical description
of the non-\\perturbative interactions between charged
 particles and soft photons. At zero temperature, it provides 
the correct structure of the fermion propagator near the mass-shell 
\cite{Bogoliubov}.  At finite temperature, 
the Bloch-Nordsieck approximation has been used,
 by Weldon, to verify the cancellation of the infrared divergences
in the production rate for soft photons \cite{Weldon94}.
The remarkable structure of the ``hard thermal loops'' (HTL) \cite{BP90}
emerges from similar kinematical approximations, as clearly
 emphasized in the kinetic derivation of the HTL's \cite{BI94,BIO96}.	
More recently, a similar approximation has been used in Refs.
 \cite{prl,prd} to eliminate the infrared divergences in 
the computation of the fermion damping 
rate \cite{Pisarski89,Lebedev90,Pisarski93,all}.
As shown in Ref.~\cite{prd}, this calculation requires the resummation
of an infinite class of multi-loop Feynman graphs of the type 
shown in Fig.~\ref{Nloop}.
These are the same diagrams as those of the quenched approximation
(i.e., all fermion loops are ignored),
except for the fact that the photon lines include the hard thermal loop
correction.

Throughout this work, we shall be mainly interested in the {\it leading}
contribution of such diagrams to the propagator of a hard fermion
($p\simge T$) near its mass-shell ($p_0\sim p$). 
As shown in Ref.~\cite{prd}, this leading contribution
can be estimated in the Bloch-Nordsieck approximation, that is,
with the following simplified Feynman rules: (i) the fermion propagator 
\beq\label{G0}
G_0(p-q)\,=\,\frac{1}{(p^0-q^0)-{\bf v}\cdot ({\bf p-q})}\,;\eeq
(ii) the photon-fermion vertex $\Gamma^\mu=v^\mu$, and (iii)
 the HTL photon propagator   ${}^*D_{\mu\nu}(q)$.
Here, $p^\mu=(p_0,{\bf p})$ is the  external hard momentum,
with $p_0\simeq p$, ${\bf v}={\bf p}/p$ is the corresponding velocity,
 $v^\mu\equiv (1,{\bf v})$, and $q^\mu=
(q_0,{\bf q})$ is a linear combination of the soft momenta of the
internal photons.
The above Feynman rules govern the interactions
between any kind of {\it hard} charged (or colored) particles,
irrespective of their spin, and {\it soft} gauge fields,
to leading order in an expansion in powers of
the soft momenta \cite{Weinberg} (see also
Appendix C below for a discussion of the non-Abelian case).
They lead to simplifications by ignoring those degrees of freedom
 --- in this case, spin and negative-energy states ---
which play no dynamical role in the kinematical regime of interest.

The imaginary part of the fermion self-energy computed with these rules
exhibits infrared divergences near the mass-shell,
 to all orders in perturbation theory.
For instance, in the one-loop approximation we have:
\beq\label{1loop} {\rm Im}\,\Sigma_R^{(2)}(\omega \simeq p)\,\simeq\,-
\alpha T\ln \frac{\omega_p}{|\omega-p|},
\eeq
where $\alpha \equiv g^2/4\pi$, $\omega_p=gT/3$ (the plasma
frequency), and the approximate equality means that only the singular
term has been preserved. 
For two or more photon loops, the mass-shell divergences are power-like
\cite{prd}. 
Such divergences  prevent us from computing
the mass-shell structure of the charged particles, and
in particular from obtaining the fermion lifetime in perturbation theory.

Note, however,  that  no infrared divergences are encountered when
the perturbation theory is carried out directly 
in the {\it time} representation: the inverse of the time acts 
then effectively as an infrared cutoff. For instance,
the one-loop correction to the retarded propagator $S_R(t,{\bf p})$
 at large times is given by:
\beq\label{LT}
\delta S_R^{(2)}(t,{\bf p})\,\simeq\,-it\int_{0}^t {\rm d}t' 
\, {\rm e}^{ipt'}\,\Sigma_R^{(2)}(t',{\bf p})\,.\eeq
This expression
 is well defined although the limit $t\to \infty$ of the integral over  $t'$
(which is precisely the on-shell self-energy 
$\Sigma_R^{(2)}(\omega = p)$) does not exist.
We actually have \cite{prd}
\beq\label{SRT1}
\Sigma_R^{(2)}(t, {\bf p})\,\simeq\,
-i\alpha T \,
\frac{{\rm e}^{-ipt}}{t}\,\,\,\qquad {\rm for}\,\,\,\,\,\,\,
t\gg \frac {1}{\omega_p}\,,\eeq
and therefore
\beq\label{LT2}
\delta S_R^{(2)}(t,{\bf p})\,\simeq\,-
\alpha T t\int_{1/\omega_p}^t
 \frac{{\rm d}t'}{t'}\,=\,-\alpha Tt \ln (\omega_p t).\eeq
As shown in Refs. \cite{prl,prd}, this correction exponentiates
in an all-order calculation:
\beq\label{DLT0}
S_R(t,{\bf p})\,\propto\, {\rm exp}\Bigl( -\alpha Tt \ln \omega_p
t\Bigr)\,\,\,\qquad {\rm for}\,\,\,\,\,\,\,
t\gg \frac {1}{\omega_p}\,.\eeq

Note, however, that the approximations used in Refs. \cite{prl,prd},
and which lead to eq.~(\ref{DLT0}),
are reliable only for computing the {\it leading}
large-time behaviour displayed in eq.~(\ref{DLT0}).
(These approximations involve, aside from the
Bloch-Nordsieck approximation, also a restriction to the static photon
mode.) The subleading term --- i.e., the constant
term under the logarithm --- could not be
obtained in this way, and the issue of its gauge (in)dependence
remained an entirely open problem.

In this paper, we improve the accuracy of our previous calculation
by also including the non-static photon modes within
the Bloch-Nordsieck calculation. This is sufficient
to fix the term of order $g^2T$ in the exponential
(\ref{DLT0}). As we shall see, this term, which receives contributions
from both the electric and the magnetic sectors, becomes gauge-fixing
independent when an infrared regulator is introduced in the gauge 
sector. The final result, to be derived below, is
\beq\label{DLT}
S_R(t\gg 1/\omega_p)\,\propto\, {\rm exp}\left\{ -\alpha Tt\Bigl(
 \ln(\omega_p t)\,+\,0.12652...\,+\,{\cal O}(g)\Bigr)\right\}.\eeq
This result applies to a massless fermion with momentum
$p\sim T$ or larger. The extension
to a massive ($m\gg T$) test particle is straightforward.
The case of a  soft fermion ($p\sim gT$), on the other hand,
 requires the full machinery
of the HTL-resummation \cite{BP90}, and will be not addressed here
(see Ref. \cite{prd} for the leading order result in this case).

In order to derive eq.~(\ref{DLT}), we shall use a
finite-temperature extension of the Bloch-Nordsieck (BN) model,
to be introduced in section 2.
Formally, our construction is a straightforward generalization
of the corresponding model at zero-temperature, as described
for example in Ref. \cite{Bogoliubov}. However, unlike
what happens at zero-temperature, at finite-temperature, the BN model cannot
be solved in closed form (see also \cite{prd}). 
The technical difficulty comes from the thermal boundary conditions
to be imposed on the BN propagator,
and more specifically from the thermal occupation 
factors for the hard fermion. However, as
 it will be explained in Sections 3 and 4, this
problem can be overcome, within the desired accuracy,
 and this eventually yields the
large-time behaviour indicated in eq.~(\ref{DLT}).
The independence of this result with respect to the choice
of the gauge is further analyzed in section 5.
Finally, in section 6, we consider an extension of the thermal BN model to QCD.
Because of the mutual interactions  of the soft gluons, the non-Abelian
model cannot be solved analytically. 
Our main result here is an expression of the retarded propagator
of a hard  quark or gluon in the form of a
functional integral over three-dimensional Euclidean gauge fields.
This representation, which is reminiscent of the dimensional reduction
sometimes performed in the computation of { static} thermal
correlation functions \cite{Braaten94,Kajantie94},
is well adapted to numerical calculations on a lattice.

\section{The Bloch-Nordsieck propagator}
\setcounter{equation}{0}

As mentioned in the Introduction, we are interested in the
large time decay of the propagator of a hard fermion
moving through a QED plasma at very high temperature $T$ :
$T\gg m_e$, where $m_e$ is the electron mass in the vacuum.
This fermion can be either a thermal electron, with typical
momentum $p\sim T$ and ultrarelativistic dispersion relation
$E_p=p$ (we neglect the electron mass relative to $T$),
or a (generally massive) test charged particle, with
 three-momentum $p\simge T$ and  dispersion relation
 $E_p=\sqrt{p^2+m^2}$. By test particle, we mean a
particle which is distinguishable from the plasma particles,
 and is therefore not part of the thermal bath.
The general formalism below will be developed 
for a thermal particle. We shall indicate later how one can
derive from it the simpler case of the test particle.
Also, we shall write the general formulae for a
massless fermion. The corresponding formulae
for a massive test particle will be presented only briefly.

\begin{figure}
\protect \epsfxsize=16.cm{\centerline{\epsfbox{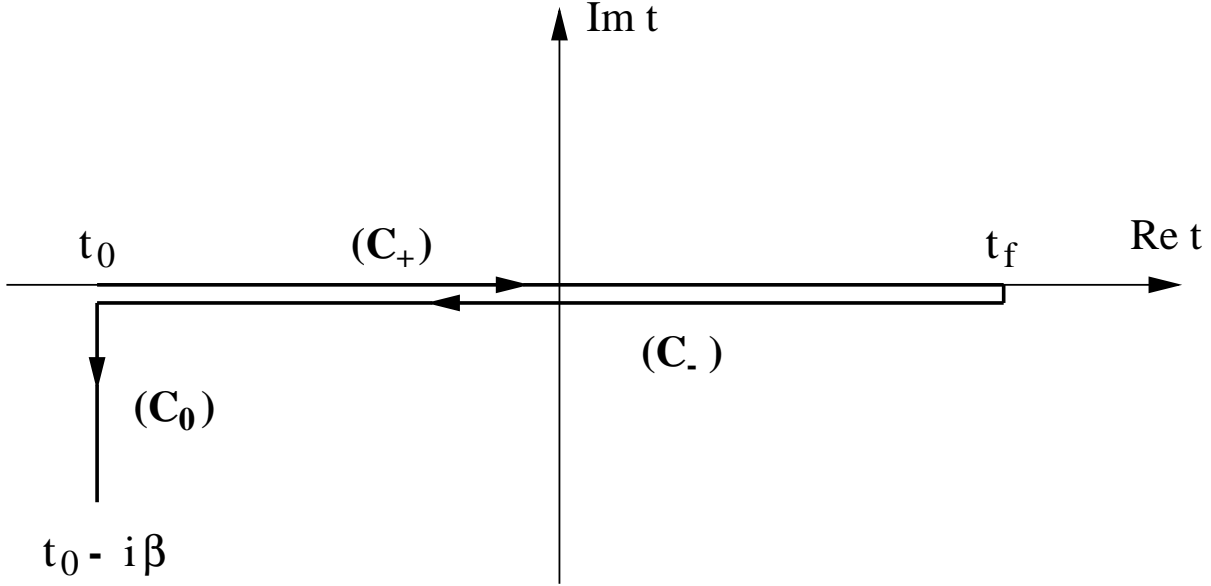}}}
	 \caption{Complex-time contour for the evaluation of
the thermal expectation values: $C=C_+\cup C_-\cup C_0$. 
On $C_+$, $t$ takes all the
real values between $t_0$ to $t_f$ (eventually, we let
$t_f\to\infty$). On  $C_-$,  $t\,\to\,t-i0_+$, where $t$ runs backward from
$t_f$ to $t_0$. Finally, on $C_0$, $t=t_0-i\tau$, with $0<\tau\le \beta$.}
\label{CONT}
\end{figure}
We are eventually interested in the retarded propagator,
\beq\label{SR0}
iS_R(x-y)\,\equiv\,\theta(x_0-y_0)\,\langle\{\psi(x),\,
\bar\psi(y)\}\rangle,\eeq
where the curly braces denote the anticommutator
of the fermion field operators, and the angular
braces, the thermal expectation value. However, to
calculate this propagator at finite temperature,
it is  convenient to consider first the time-ordered (or Feynman)
propagator,
\beq\label{SF}
iS(x-y)\,\equiv\,\langle {\rm T} \psi(x)
\bar\psi(y)\rangle\,=\,\theta(x_0-y_0)\,S^>(x-y)-\,
\theta(y_0-x_0)\,S^<(x-y)\,,\eeq
and to observe that the two two-point functions
\beq\label{S><}
S^>(x-y)\,\equiv\,\langle\psi(x) \bar\psi(y)\rangle\,,\qquad\,\,
S^<(x-y)\,\equiv\,\langle \bar\psi(y)\psi(x)\rangle\,,\eeq
are analytic functions of their time arguments. These functions
 can be first computed in the imaginary-time formalism
 \cite{MLB96}, and then continued to the real-time axis.
Then, the retarded propagator (\ref{SR0}) can be obtained as
\beq\label{SR02}
iS_R(x-y)\,=\,\theta(x_0-y_0)\,\Bigl(S^>(x-y)\,+\,
S^<(x-y)\Bigr)\,.\eeq
This is a method that we have used in Ref. \cite{prd}, and we shall
use it again in Appendix A where we
compute the propagator of a hard thermal fermion.

An alternative formalism, which permits a direct evaluation
of real-time Green's functions, is based on the use
of an oriented contour $C$ 
in the complex time plane, as shown in Fig.~\ref{CONT} \cite{MLB96}.
We define the {\it contour-ordered} propagator
\beq\label{SC1}
iS(x-y)\,\equiv\,\langle {\rm T}_C \psi(x)
\bar\psi(y)\rangle\,=\,\theta_C(x_0,y_0)\,S^>(x-y)\,-\,
\theta_C(y_0,x_0)\,S^<(x-y),\eeq
where the time variables $x_0$ and $y_0$ lie on $C$,
and ${\rm T}_C$ and $\theta_C$ denote respectively
the contour-ordering operator and the contour theta function.
(If one gives a parametric representation of the path,
$t=z(u)$, with $u$ real and monotonically increasing,
then path ordering corresponds to the ordering in $u$,
and $\theta_C(t_1,t_2)=\theta(u_1-u_2)$.)
The contour propagator (\ref{SC1}) satisfies the Kubo-Martin-Schwinger
(KMS) boundary condition \cite{MLB96}
\beq\label{KMS}
S(t_0-y_0)\,=\,-\,S(t_0-y_0-i\beta).\eeq
(In this equation, and often below, we omit the spatial coordinates,
for simplicity.) It
 can be given the following spectral representation  \cite{MLB96}:
\beq\label{specS}
iS(x-y)&=&\int\frac{{\rm d}^4p}{(2\pi)^4}\,{\rm e}^{-ip\cdot(x-y)}\,
\acute\rho(p)\Bigl[\theta_C(x_0,y_0)-n(p_0)\Bigr]
\,,\eeq
where $\acute\rho(p)$ is the fermion spectral density and
$n(p_0) = 1/\bigl({\rm exp}(\beta p_0)+1\bigr)$. Note that,
once the spectral density is known, the retarded propagator (\ref{SR0})
can be obtained as:
\beq\label{specSR}
iS_R(x-y)
\,=\,\theta(x_0-y_0)
\int\frac{{\rm d}^4p}{(2\pi)^4}\,{\rm e}^{-ip\cdot(x-y)}\,\acute\rho(p)
\,.\,\eeq
In fact, in cases where we shall use the contour method below,
the relation between $S$ and $S_R$ will be even simpler.
Indeed, in these cases --- that of a test particle,
and that of a very energetic thermal particle, with $p\gg T $ ---
the statistical factor
$n(p_0)$ can simply be ignored, so that eqs.~(\ref{specS})
and (\ref{specSR}) become identical for $x_0$ and $y_0$ real.

The large time behaviour of the fermion propagator
is governed by the interactions of the fermion 
with soft thermal photons. These
 can be analyzed in the BN approximation. The propagator
 has then the following functional integral
representation \cite{Bogoliubov} (see also Ref. \cite{prd})
\beq\label{FunctionalS}
S(x-y)= Z^{-1}\int  {\cal D} A \,G(x,y|A)\,{\rm e}^{iS_C[A]},\eeq
where  $G(x,y|A)$ is the solution of the equation:
\beq\label{SA}
i (v\cdot D_x)  G(x,y|A)=\delta_C (x,y),\eeq
where $D_\mu=\del_\mu +igA_\mu$, $v^\mu \equiv (1,{\bf v})$,
$|{\bf v}|=1$, $\delta_C(x,y)$ is the contour delta function \cite{MLB96},
and $S_C[A]$ is the effective action for soft photons in the
 hard thermal loop (HTL) approximation \cite{BP90,BIO96,MLB96}:
\beq\label{SEFF} 
S_C[A]&=&\int_C {\rm d}^4x\left\{-\frac{1}{4}\,F_{\mu\nu}F^{\mu\nu}
-\frac{1}{2\lambda}(\del\cdot A)^2\right\}
\,+\,\int_C {\rm d}^4x\int_C {\rm d}^4y\,\,\frac{1}{2}\,A^\mu(x)
\Pi_{\mu\nu}(x,y)A^\nu(y)\nonumber\\
&\equiv&\int_C {\rm d}^4x\int_C {\rm d}^4y\,\,\frac{1}{2}\,A^\mu(x)
D_{\mu\nu}^{-1} (x-y) A^\nu(y).\eeq
We have written this equation in the covariant
gauge with parameter $\lambda$. The Coulomb gauge $\bfgrad \cdot {\bf A}=0$
will also be used in what follows. 
$\Pi_{\mu\nu}$ is the photon polarization tensor
in the HTL-approximation \cite{Klimov81,BP90}.

The gauge fields to be integrated over
in eq.~(\ref{FunctionalS}) satisfy the periodicity condition
$A_\mu(t_0,{\bf x})=A_\mu(t_0-i\beta,{\bf x})$. 
Correspondingly, the photon  contour propagator satisfies
\beq D_{\mu\nu}(t_0-y_0)\,=\,
D_{\mu\nu}(t_0-y_0-i\beta),\eeq
and can be given the following spectral representation:
\beq\label{specD}
D_{\mu\nu}(x-y)=-i\int\frac{{\rm d}^4q}{(2\pi)^4}\,{\rm e}^{-iq\cdot(x-y)}
\rho_{\mu\nu}(q)\Bigl[\theta_C(x_0,y_0)+N(q_0)\Bigr]\,,\eeq
where $\rho_{\mu\nu}(q)$ is the photon spectral density in the HTL 
approximation \cite{Pisarski89a}
(we follow here the notations in Ref. \cite{prd}),
 and $N(q_0) = 1/\bigl({\rm exp}(\beta q_0)-1\bigr)$
is the Bose-Einstein statistical factor.

Eq.~(\ref{SA}) defines the BN propagator of a charged
particle in a classical background
field $A_\mu(x)$. The vector ${\bf v}$ in this equation is to be
 identified with the  particle velocity.
Then, the Feynman rules generated by
the functional integral (\ref{FunctionalS}) coincide with
those given in the introduction.
Since, in the BN model, ${\bf v}$ is a fixed parameter, 
the underlying physical approximation is the lack of fermion recoil.
This approximation is justified when the
momentum transfer from the background field is small,
at most of order $\sim gT$.
(The energy-momentum scale $gT$ is set 
by the polarisation tensor $\Pi_{\mu\nu}$; see section 3 below.)

In order to satisfy the KMS condition
(\ref{KMS}), the BN equation  (\ref{SA}) 
has to be solved for antiperiodic boundary conditions:
\beq\label{BC} G(t_0,y_0|A)=- \,G(t_0-i\beta,y_0|A),\eeq
and similarly for $y_0$. These conditions complicate
the resolution of the thermal BN model in general.
There is a  simple case, however, where this complication is absent,
namely the case of a test particle. For this case,
the thermal BN model can be exactly solved, as we discuss now.

\subsection{The test particle}

The propagator of a test particle has only one analytic
component, namely $S^>(x-y)$; $S^<$ vanishes since the
thermal bath acts like the vacuum for the field operators
of the test particle. Therefore (cf. eqs.~(\ref{SR0})--(\ref{SR02})),
$S(x-y)=S_R(x-y)=-i\theta(x_0-y_0)S^>(x-y)$ and the KMS conditions
(\ref{KMS}) do not apply. In the BN approximation,
$S(x,y)$ is still given by eq.~(\ref{FunctionalS}), but now
$G(x,y|A)$ obeys retarded conditions in real time,
and not the conditions (\ref{BC}). The solution to
eq.~(\ref{SA}) is then obtained in closed form
\cite{Bogoliubov} (see also Ref. \cite{prd}) :
\beq\label{GR}
G_R(x,y|A)&=&-i\,\theta (x^0-y^0)\,\delta^{(3)}
\left({{\bf x}}-{{\bf y}}-{{\bf v}}(x^0-y^0)
\right )U(x,y)\nonumber\\
&=&-i\,\int_0^\infty {\rm d}t\,\delta^{(4)}(x-y-vt)\,U(x,x-vt),\eeq
where the time variables $x_0$ and $y_0$ are real.
The three-dimensional delta function describes straightline propagation
with velocity ${\bf v}$. The background gauge field only
contributes a phase factor:
\beq\label{U}
U(x,x-vt)\,\equiv\,\,\exp\left\{ -ig\int_0^t {\rm d}s \, v\cdot A(x-v(t - s))
\right\}.\eeq
In momentum space, the free ($A_\mu=0$) retarded propagator reads
 \beq\label{G0R}G_R^{\,0}(\omega, {\bf p})
\,=\,\frac{1}{\omega-{\bf v\cdot p} +i\eta}\,,\eeq
corresponding to the following free spectral density:
\beq\label{newrho}
\acute\rho_0(\omega, {\bf p})\equiv -\,2{\rm Im}\,G_R^{\,0}(\omega, {\bf p})
 \,=\,2\pi\delta(\omega-{\bf v}\cdot {\bf p}).\eeq
According to these equations, the fermion mass-shell in the BN model
corresponds to $\omega=E_p\equiv {\bf v}\cdot {\bf p}$. Since
 ${\bf v}$ is to be identified with the velocity
${\bf p}/p$ of the massless fermion, the 
free mass-shell is at $\omega=p$, as it should.

In order to perform  the functional integration (\ref{FunctionalS}),
we first rewrite the parallel transporter (\ref{U}) as follows:
\beq \label{jz}
U(x,x-vt)&=&\exp\left\{ -i\int{\rm d}^4z\, j_\mu(z) A^\mu(z)
\right\},\nonumber\\
 j_\mu(z)&\equiv & g v_\mu\int_0^t {\rm d}s \, 
\delta^{(4)}(z-x+v(t-s)).\eeq
Then, a straightforward calculation yields:
\beq\label{SRT}
S_R(t,{\bf p})&=&-i\theta(t) {\rm e}^{-it({\bf v\cdot p})}\,\tilde\Delta(t),\eeq
with 
\beq\label{Delta1}
\tilde\Delta(t) &=&{\rm exp}\left \{-\,\frac{i}{2}\,
\int_{C_+} {\rm d}^4x\int_{C_+} {\rm d}^4y\,j^\mu(x)
\,D_{\mu\nu}(x-y)\,j^\nu(y)\right\},\eeq
where the time integrals run on $C_+$ only, in accordance
with eq.~(\ref{jz}).

By using eqs.~(\ref{jz}) and (\ref{specD}), and after a simple calculation,
we can rewrite $\tilde\Delta(t)$ in eq.~(\ref{Delta1}) as follows:
\beq\label{Delta20}
\tilde\Delta(t)=
 {\rm exp}\left\{-\,\frac{g^2}{2}\int\frac{{\rm d}^4q}{(2\pi)^4}
\int_0^t {\rm d}s_1 \int_0^t {\rm d}s_2\,{\rm e}^{-i(q\cdot v)(s_1-s_2)}\,
\tilde \rho(q)\Bigl[\theta(s_1-s_2)+N(q_0)\Bigr] \right \},\,\,\eeq
with $\tilde \rho(q)\equiv v^\mu\rho_{\mu\nu}(q)v^\nu$.
Note that eq.~(\ref{Delta20}) could have been obtained 
from the corresponding expression in the vacuum (see Ref. \cite{Bogoliubov})
by simply replacing in the latter the bare photon propagator by the
corresponding thermal propagator for a soft photon.
Converserly, the zero-temperature BN propagator can be obtained
from eq.~(\ref{Delta20}) by substituting $N(q_0)\,\to\,
-\theta(-q_0)$ and replacing $\rho_{\mu\nu}$ with the free 
 photon spectral density. 

After performing the $s_1$ and $s_2$ integrations,
and also using the parity property $\tilde\rho(-q)=-\tilde\rho(q)$,
we finally cast eq.~(\ref{Delta20}) into the form :
\beq \tilde\Delta(t)&=& {\rm exp}\{it\Phi(t)\}\,\Delta(t),\eeq
where
\beq\label{PHI}
\Phi(t)\equiv g^2 \int \frac{{\rm d}^4q}{(2\pi)^4}
 \,\,\frac{\tilde\rho(q)}{2(v\cdot q)}\left[1\,-\,
\frac{\sin t(v\cdot q)}{t(v\cdot q)}\right],\eeq
and
\beq\label{Delta2}
\Delta(t)\equiv  {\rm exp}\left \{-g^2
\int \frac{{\rm d}^4q}{(2\pi)^4} \,\,\tilde\rho(q)\,N(q_0)\,\,
\frac{1-  {\rm cos}\,t({v\cdot q})}{(v\cdot q)^2}\,
\right\}.\eeq

In the derivation of the above formulae, 
eqs.~(\ref{Delta20})--(\ref{Delta2}),
there was no explicit restriction on the photon momenta $q^\mu$.
Since the BN model can only be trusted for soft photons,
we need to verify that the large time behaviour of the fermion propagator,
as given by eqs.~(\ref{SRT})--(\ref{Delta2}), is indeed controlled
by soft  momenta, $q\ll T$.

In fact, the momentum integrals in eqs.~(\ref{PHI})--(\ref{Delta2})
 contain ultraviolet divergences
coming from their zero-temperature contributions.
There is a linear UV divergence in  $\Phi(t)$, and a
logarithmic divergence in $\Delta(t)$. These divergences
can be absorbed respectively, by mass and field-strength renormalizations
\cite{Bogoliubov}. However, the finite part of the phase $\Phi(t)$
remains dominated by hard momenta  contributions, and therefore
is not consistently determined by the present approximation. 
Since  $\Phi(t)$ does not enter the calculation of the lifetime, 
we shall ignore it in what follows.

The damping effects are entirely described by the function
$\Delta(t)$, eq.~(\ref{Delta2}), which
extends our previous result \cite{prl,prd}
by including the effects of the non static ($q_0\ne 0$)
electric and magnetic field fluctuations. At $T=0$,
 $\Delta_{T=0}\propto {\rm exp}(-g^2 \ln \Lambda t)$,
where $\Lambda$ is the upper momentum cutoff \cite{Bogoliubov}.
After UV renormalization, the cutoff $\Lambda$ is replaced
by the physical electron mass, thus yielding
\beq
|S(t)|\sim \left(mt\right)^{(3-\lambda)\,\frac{\alpha}{2\pi}},\eeq
in the covariant gauge with gauge-fixing parameter $\lambda$
(see eq.~(\ref{SRT0})). Such a gauge-dependent, polynomial
dependence on time merely reflects the renormalization of
the wave-function of the fermion due to its coupling to soft
virtual photons. This is not to be interpreted as a damping phenomenon.
The mechanism which takes place at high temperature and which
eventually gives rise to the damping of the test particle
excitation is the exchange of soft photons between the test
fermion and the thermal charged particles.
We shall verify in section 3 that, for sufficiently large times 
$t\gg 1/gT$,  such a collision involves dominantly
soft photon momenta $q\simle gT$.

For a fixed large time $t$, the function 
\beq\label{Asymp}
f(t,{v\cdot q})&\equiv &\frac{1-{\rm cos}\,t(v\cdot q)}{(v\cdot q)^2}\,,
\eeq
in eq.~(\ref{Delta2}) is strongly peaked around ${v\cdot q}\equiv
q_0-{\bf v\cdot q}=0$, 
with a width $\sim 1/t$. In the limit $t\to\infty$,
$f(t,{v\cdot q})\,\to\,\pi t \delta (v\cdot q)$.
In the absence of infrared complications, we could use this
limit to obtain the large time behaviour of  $\Delta(t)$. This
procedure would then yield
 $\Delta(t\to \infty)\,\sim\,{\rm e}^{-\gamma t}$, with:
\beq\label{naive}
\gamma &\equiv&\pi g^2\int \frac{{\rm d}^4q}{(2\pi)^4} \,\tilde\rho(q)\,N(q_0)\,
\delta (v\cdot q)\,.\eeq
We recognize in eq.~(\ref{naive})
the one-loop damping rate $\gamma = -\, {\rm Im}\,\Sigma^{(2)}(\omega =p)$
\cite{Pisarski93,all}, which we know, however, to be
 infrared divergent (cf. eq.~(\ref{1loop})).
Thus, in studying the large-time behaviour of eq.~(\ref{Delta2}),
one should keep the time finite when performing the momentum integral.
As already mentioned after eq.~(\ref{LT}), 
the inverse time plays the role of an infrared cutoff.
This will become explicit in section 3 below.

Eqs.~(\ref{SRT})--(\ref{Delta2}) generalize trivially
to a test particle with mass $m$. The mass-shell is shifted to
 $E_p\equiv  {\bf v}\cdot {\bf p}+m(1-{\bf v}^2)^{1/2}$,
(which, since ${\bf v}={\bf p}/E_p$,
 corresponds indeed to $E_p=\sqrt{{\bf p}^2+m^2}$),
and the retarded propagator has the form (\ref{SRT}):
\beq\label{SRTM}
|S_R(t,{\bf p})|&=&\theta(t)\,\Delta_v(t),\eeq
where $\Delta_v(t)$ is the function (\ref{Delta2}) with, however,
$|{\bf v}| < 1$. 

\subsection{The thermal fermion}

The case of a thermal electron
with momentum $p\sim T$ is
physically more interesting, since this is a typical quasiparticle
of the  plasma. Technically, however, this is more involved, since
the KMS boundary conditions (\ref{BC}) must be taken into account.

To appreciate the difficulty, consider the
 free contour propagator, as obtained by replacing
 $\acute\rho(p)$ with $\acute\rho_0(p)=
2\pi\delta(\omega-{\bf v}\cdot {\bf p})$ in eq.~(\ref{specS}):
\beq\label{GOtau}
G_0(t-t',{\bf p})\,=\,-i{\rm e}^{- i({\bf v\cdot p})
(t-t')}\Bigl[\theta_C(t,t')(1-n_p)
\,-\,\theta_C(t',t)n_p\Bigr],\eeq
where $n_p\equiv n({\bf v\cdot p})$.
By using this propagator, we can solve  the BN equation 
(\ref{SA}) as a series  in powers of $gA_\mu$.
To this aim, one can first transform eq.~(\ref{SA})
into an integral equation,
\beq
G(x,y|A)\,=\,G_0(x-y)\,+\,g\int_C{\rm d}^4z \,G_0(x-z) \,v\cdot A(z)\,
G(z,y|A)\,.\eeq
Then, by iteratively solving this equation, one
 generates the perturbation series for $G(x,y|A)$.
However, in contrast to what happens for the retarded propagator
(\ref{GR}), the resulting series for the contour propagator $G(x,y|A)$
does {\it not} exponentiate \cite{prd}.
The exponentiation of the perturbative series for $G_R(x,y|A)$
is related to the fact that the  retarded free propagator,
\beq\label{GR0t}
G_R^{\,0}(t,{\bf p})\,=\,-i\theta(t){\rm e}^{- i({\bf v\cdot p})t}\,,\eeq
satisfies the simple multiplication law
$G_R^{\,0}(t,{\bf p}_1)\,G_R^{\,0}(t,{\bf p}_2)
\,=\,-G_R^{\,0}(t,{\bf p}_1+{\bf p}_2)$.
The contour propagator $G_0(t,{\bf p})$ does not enjoy this property, 
because of the presence of the statistical factors
 in eq.~(\ref{GOtau}).

This argument suggests that the contour BN propagator may 
exponentiate  whenever the fermion occupation numbers play no dynamical role.
This is  what happened for the test particle in the previous
subsection, and, more generally, this will also happen
for a thermalized fermion with very high momentum,
$p\gg T$, whose thermal occupation number is exponentially small:
$ n(p)\,\simeq\,{\rm e}^{-\beta p}\,\ll\,1$.
In fact, when $n_p\to 0$, the free contour propagator (\ref{GOtau})
reduces to the retarded function (\ref{GR0t}) (for real time variables).
It is then easy to verify that the previous solution
of the BN model, as given by eqs.~(\ref{SRT})--(\ref{Delta2}),
also applies to such a very energetic thermal particle,
up to corrections which are  exponentially small when $p\gg T$.
In particular, the case of the test particle is formally
recovered as the limit $p/T\to \infty$.

What is less obvious
is that the same solution holds also for a typical thermal
fermion, with momentum $p\sim T$. More precisely,
as will be verified in Appendix A, 
the thermal fermion  propagator decays according to the same
law as above, that is,
\beq\label{SR1}
|S_R(t,{\bf p})|&\propto&\Delta(t),\eeq
(with  $\Delta(t)$ as defined in eq.~(\ref{Delta2})),
up to corrections of order $q/T \simle g$. 
Physically, this reflects the fact (which has been
already mentioned at several places, and will be verified
in the next section) that the fermion decay at large times, $t\gg 1/gT$,
is determined by its interactions with soft photons, with momenta
$q\simle gT$. Such interactions do not
significantly change the electron momentum, so that the
associated thermal occupation factors play no dynamical role.

\section{Large time behaviour}
\setcounter{equation}{0}

We are now in a position to study the large-time behaviour of the
fermion propagator, as described
by  the function $\Delta(t)$,  eq.~(\ref{Delta2}).
We shall verify below that the relevant energy scale is 
hidden in the photon spectral density $\rho_{\mu\nu}(q)$,
and is of the order $gT$.
Therefore, ``large times'' means times larger than $1/gT$.

For the computation below, we shall use the photon
 spectral density in the Coulomb gauge:
\beq\label{Coul}
 \tilde \rho(q_0,{\bf q})\,=\,\rho_l(q_0,q)\,+
\,\left(1- ({\bf v}\cdot \hat{\bf q})^2\right)\,\rho_t(q_0,q).\eeq
(The issue of the gauge dependence will be addressed in the next section.)
The two pieces $\rho_l(q_0,q)$ and $\rho_t(q_0,q)$ of the spectral density
correspond respectively to longitudinal and transverse photons,
which are renormalized  differently by plasma effects \cite{BIO96,MLB96}.
They have the following structure 
(with $s=l$ or $t$):
\beq \label{RRHO}
\rho_s(q_0,q)
= 2\pi\epsilon(q_0)\,z_s(q)\,\delta(q_0^2 -\omega_s^2(q))
 +\beta_s (q_0,q) \theta (q^2- q_0^2),\eeq
and involve delta functions associated to plasma waves at time-like
 momenta ($q_0^2={\omega_s^2(q)}>q^2$), and smooth contributions
$\beta_l$ and $\beta_t$ at $q_0^2<q^2$ arising from Landau damping.
For given $q_0$ and $q$, the energy-momentum scale in
eq.~(\ref{RRHO}) is set by the plasma frequency $\omega_p\equiv
gT/3$, for both the on-shell and the off-shell spectral densities
(see \cite{Pisarski89a,BIO96,MLB96} for more details).

For generic times, both pieces in eq.~(\ref{RRHO})
contribute to  eq.~(\ref{Delta2}):\\
i) That involving $ \delta(q_0^2 -\omega_s^2(q))$
describes the emission or the absorption of an on-shell plasmon.
By kinematics, this is only possible if the fermion is sufficiently
off-shell, $|\omega-p|\simge gT$: indeed, the  plasmons propagate
as massive particles, with (momentum dependent) thermal
masses of order $gT$ \cite{Klimov81,BIO96,MLB96}.\\
ii) The contributions involving $\beta_l$ and $\beta_t$ describe
{\it collisional} damping, where the fermion exchanges
a virtual photon with the other charged particles of the plasma.
Such processes have no kinematical restrictions, and they are
the only one to contribute at very large times $t\gg 1/gT$.

To study the large-time behaviour, we restrict therefore ourselves to
collisional processes, i.e., retain only $\beta_l$ and $\beta_t$
in the photon spectral functions. From  perturbation theory,
we know that the infrared complications are related 
to the singular behaviour of the magnetic spectral density
as $q_0\ll q\to 0$ \cite{prd} :
\beq\label{rhot0}
\frac{1}{q_0}\,\beta_t(q_0\ll q)\,\simeq\, 
\frac{3\pi}{2}\, \frac{\omega_p^2 \, q}
{q^6\,+\, (3\pi \omega_p^2 q_0/4)^2}\,\,\to\,\,
\frac{2\pi}{q^2}\,\delta(q_0)\,\,\,\,\,\, {\rm as} \,\,\,\, q\to 0.\,\,\eeq
To isolate this singular behaviour, we write
\beq\label{sept}
\frac{1}{q_0}\,\beta_t(q_0,q)\,\equiv\,
2\pi\delta(q_0)\left(\frac{1}{q^2}\,-\,\frac{1}{q^2+\omega_p^2}\right)\,+\,
\frac{1}{q_0}\,\nu_t(q_0,q).\eeq
A contribution $\propto 1/(q^2+\omega_p^2)$ has been subtracted from 
the singular piece
--- and implicitly included in $\nu_t(q_0,q)$ --- to avoid spurious ultraviolet 
divergences: written as they stand, both terms in the
 r.h.s. of eq.~(\ref{sept}) give UV-finite contributions. 
Note that by neglecting the regular piece  $\nu_t(q_0,q)$ in the 
right hand side of eq.~(\ref{sept}), one recovers our previous
result in Refs. \cite{prl,prd} (as also expressed in eqs.~(\ref{FIR})
and (\ref{qLT}) below). 

With (\ref{sept}), the integral in eq.~(\ref{Delta2}) may be separated
into two pieces: 
\beq\label{Freg}
F_{reg}(t)\,\equiv\, \int \frac{{\rm d}^3q}{(2\pi)^3}
 \int_{-q}^q \frac{{\rm d}q_0}{2\pi q_0}\left(
\beta_l(q_0,q) -\cos^2\theta\,\beta_t(q_0,q)
+ \nu_t(q_0,q)\right)\frac{1-  {\rm cos}\,t(v\cdot q)}{(v\cdot q)^2}\,,\eeq
and
\beq\label{FIR}
F_{IR}(t)\,\equiv\, \int \frac{{\rm d}^3q}{(2\pi)^3}
\left(\frac{1}{q^2}\,-\,\frac{1}{q^2+\omega_p^2}\right)\,
\frac{1-  {\rm cos}\,t({\vq})}{({\vq})^2}\,.\eeq
The first piece, $F_{reg}(t)$, is infrared safe, and its large-time 
limit can be taken by replacing
$f(t,{v\cdot q})$ by $\pi t \delta (v\cdot q)$ (see
eq.~(\ref{Asymp})). This yields
\beq\label{FregLT}
F_{reg}(t)\,\equiv\, \frac{t}{4\pi}\int_0^\infty {\rm d}q \,q
 \int_{-q}^q \frac{{\rm d}q_0}{2\pi q_0}\left(
\beta_l(q_0,q) \,-\,\frac{q_0^2}{q^2}\,\beta_t(q_0,q)
+ \nu_t(q_0,q)\right)\,,\eeq
where we have used the delta function $ \delta(q_0 -q\cos \theta)$
to perform the angular integration.
The remaining integral occurs also in Ref. \cite{Pisarski93},
as part of the one-loop damping rate, and was computed there
by using sum rules plus numerical integration.
It is computed analytically in Appendix B, with the result:
\beq\label{FregF}
F_{reg}(t)\,=\,\frac{t}{8\pi}\,\ln 3\,\simeq\,\frac{t}{4\pi}\,
\times\,0.54931.\eeq
Note that this result comes entirely from the electric piece $\beta_l(q_0,q)$:
 the two magnetic pieces, $(q_0^2/q^2)\,\beta_t(q_0,q)$ and 
$\nu_t(q_0,q)$, happen to cancel each other in the final result. 
This is purely accidental, consequence
of our specific choice for the  substracted term $ 1/(q^2+\omega_p^2)$
in eq.~(\ref{sept}).

The second piece, $F_{IR}(t)$, contains the potentially singular
magnetic contribution, so that we should take the large time limit
only {\it after} performing the integral over ${\bf q}$.
This has been done in Ref. \cite{prd}, with the following result
($\gamma_E$ is the Euler constant):
\beq\label{qLT}
F_{IR}(t)\,=\, \frac{t}{4\pi}\,\Bigl(\ln \omega_p t + (\gamma_E-1) +{\cal O}
(1/\omega_p t)\Bigr).\eeq
Note that the energy scale $\omega_p$ inside the logarithm 
arises from the large momentum ($q\simge gT$) behaviour,
where the substracted term $1/(q^2+\omega_p^2)$ acts effectively
as an UV-cutoff.

The final result for the large-time propagator reads then:
\beq\label{DLTF}
\Delta (t\gg 1/\omega_p)\,\simeq\, {\rm exp}\left\{ -\alpha Tt\Bigl(
 \ln(\omega_p t)\,+\,0.12652...
\,+\,{\cal O}(g,\,1/\omega_p t)\Bigr)\right\}.\eeq
For the consistency of our approximations, it is important to observe
that this result has been obtained by integrating,
in eqs.~(\ref{FregLT}) and (\ref{FIR}), over {\it soft}
photon  momenta $q\simle gT$. While this is obvious
for eq.~(\ref{FIR}), where the two terms inside the parantheses
mutually cancel as  $q\gg gT$, it can be also verified
for eq.~(\ref{FregLT}), by using the known behaviour
of $\beta_l$ and $\beta_t$ at large photon momenta 
\cite{Pisarski89a,prd}.

For completness, let us also give the corresponding results
for a massive test particle: After reinserting the appropriate factors
of $v\equiv |{\bf v}|$ in the previous results, we get
\beq\label{DLTFM}
\Delta_v (vt\gg 1/\omega_p)\,\simeq\, {\rm exp}\left\{ -\alpha T v t\Bigl(
 \ln(\omega_p v t)\,+\,(\gamma_E-1)\,+\,C(v)\Bigr)
\right\},\eeq
where $C(v)$ is given by the following integral:
\beq\label{CV}
C(v)\,=\,\frac{1}{v^2}\int_0^\infty {\rm d}q \,q
 \int_{-vq}^{vq} \frac{{\rm d}q_0}{2\pi q_0}\left(
\beta_l(q_0,q) \,-\,\frac{q_0^2}{q^2}\,\beta_t(q_0,q)
+v^2 \nu_t(q_0,q)\right)\,.\eeq
For a very heavy particle $m\simge T$, we may consider
the non-relativistic limit $v\ll 1$ 
(this is consistent with our approximations as long as $p\simeq mv\gg gT$).
At small $v$, the leading contribution to eq.~(\ref{CV})
comes from the electric sector. The magnetic contribution involves
 a supplementary factor of $v^2$, and vanishes as $v\to 0$.
However, because of its infrared sensitivity,
the contribution of the magnetic sector is
not analytic in $v$. We evaluate this contribution in Appendix B,
where we find:
\beq \label{CVNR}
v C(v)\,=\,\frac{1}{2}\,\Bigl(1+v\ln \frac{3\pi v}{4} +\frac{v}{2}
+{\cal O}(v^2)\Bigr),\eeq
where the first term, independent of $v$, is the contribution
of the electric sector.
Together with eq.~(\ref{DLTFM}), this yields
\beq\label{DLTNR}
\Delta_v (vt\gg 1/\omega_p)\,\simeq\, {\rm exp}\left\{ -
\,\frac{\alpha T}{2}\,t\left[1\,+\,v\ln\left(\frac{3\pi}{4}\,
v^3(\omega_p t)^2\right)\,+\,v(2\gamma_E-3/2)\right]
\right\},\eeq
for $v \ll 1$. In particular, as $v\to 0$ (i.e., $m\to \infty$),
the damping is purely exponential,
with a damping rate $\gamma_0=\alpha T/2$ which coincides
with the one-loop result in Ref. \cite{Pisarski89}.
As for the $v$-dependent terms, the coefficient of the logarithm in
eq.~(\ref{DLTNR}) is the same as for the infrared-divergent piece of the
corresponding one-loop result\footnote{Actually, a different coefficient
was reported in Refs.  \cite{Pisarski89,Pisarski93}, but the
difference is apparently due to an error in the calculations there.}.

\section{Gauge-dependence}
\setcounter{equation}{0}

We show  now that the same result (\ref{DLTF}) is obtained
in general covariant gauges {provided} the large time
limit in eq.~(\ref{Delta2}) is taken
 with an infrared cutoff in the gauge sector,
in order to eliminate the contribution of the spurious
degrees of freedom.

In the covariant gauge of eq.~(\ref{SEFF}),
the photon spectral density reads
\beq\label{covrho}
\tilde\rho(q_0,{\bf q})\,=\,\left(\frac{q^2-q_0({\bf v}\cdot {\bf q})}
{q^2-q_0^2}\right)^2\rho_l(q_0,q)\,+
\,\left(1- ({\bf v}\cdot \hat{\bf q})^2\right)\rho_t(q_0,q)
\,+\,\lambda \,\rho_\lambda(q_0,{\bf q}).\eeq
The longitudinal and transverse spectral functions $\rho_l$ and
$\rho_t$ are the same as in eq.~(\ref{Coul}), and
\beq\label{lam}
\rho_\lambda(q_0,{\bf q})\equiv (q_0 - {\bf v}\cdot {\bf q})^2\,
\,2\pi \epsilon(q_0)\,\delta'(q^2),\eeq
where $\epsilon(q_0)\equiv \theta(q_0)-
\theta(-q_0)$ and $\delta'(q^2)$ is the derivative of
$\delta(q^2)$ with respect to $q^2$.

The electric and magnetic spectral functions in eq.~(\ref{covrho})
yield the same contributions to eq.~(\ref{Delta2})
as the corresponding functions in the Coulomb gauge
(cf. eqs.~(\ref{FIR}) and (\ref{FregF})).
This is obvious for the magnetic sector.
In the electric sector, the  large-time limit introduces
the delta function $\delta(v\cdot q)$ (see eq.~(\ref{Asymp})), and
the projection factor multiplying $\rho_l(q_0,q)$ in
eq.~(\ref{covrho}) is equal to one for $q_0={\bf v}\cdot {\bf q}$.
The same argument applied in the gauge sector seems to imply that
the contribution of $ \rho_\lambda(q_0,q)$ does also
vanish, because of the factor
$(q_0 - {\bf v}\cdot {\bf q})^2$ in eq.~(\ref{lam}).
However, since the spectral function $\delta'(q^2)$ has support precisely
at the integration limits $q_0=\pm q$, we should be more careful when 
taking the limit $v\cdot q \to 0$. 

The contribution of the gauge sector to $\Delta(t)$
factorizes as ${\rm exp}[{-\lambda g^2 T F_\lambda(t)}]$, where
\beq\label{Flam}
F_\lambda(t)&\equiv& \int \frac{{\rm d}^3q}{(2\pi)^3}
 \int\frac{{\rm d}q_0}{2\pi q_0}\, \rho_\lambda(q_0,{\bf q})\,
\frac{1-  {\rm cos}\,t(v\cdot q)}{(v\cdot q)^2}\,\nonumber\\
&=&\int \frac{{\rm d}^3q}{(2\pi)^3}
\int\frac{{\rm d}q_0}{ q_0}\,\,\epsilon(q_0)\,\delta'(q^2)
\Bigl[1-  {\rm cos}\,t(v\cdot q)\Bigr].\eeq
By noting that $\delta'(q^2)=(1/2q_0)({\rm d}\delta/{\rm d}q_0)$,
we can perform an integration by parts to compute the integral over
$q_0$. After also computing the angular integral, we obtain
\beq\label{Flam1}
F_\lambda(t)\,=\,\frac{1}{2\pi^2} \int_\mu^\infty
 \frac{{\rm d}q}{q^2}\left\{1\,-\,\frac{\sin 2qt}{2qt}\,-\,
\frac{1-\cos 2qt}{4}\right\}.\eeq
Although this last integral is infrared finite, we nevertheless
compute it with an infrared cutoff $\mu$.
A straightforward calculation then yields
\beq\label{Flam2}
F_\lambda(t)\,=\,\frac{1}{8\pi^2 \mu} \left[
3\,-\,2 \,\frac {\sin 2\mu t}{2\mu t}\,-\,\cos 2\mu t\right]\,-\,
\frac{t}{4\pi^2}\,{\rm si}\,(2\mu t),\eeq
where ${\rm si}(x)\equiv -\,\int_x^\infty {\rm d}z (\sin z/z)$
is the sine integral function \cite{Ober}.

If we remove the IR cutoff by letting $\mu \to 0$ at fixed $t$,
then, by using ${\rm si}(0)=-\pi/2$, we get
\beq\label{Flam11}
F_\lambda(t)\,=\,\frac{t}{8\pi}\,,\eeq
so that $\Delta(t)$ becomes (cf. eq.~(\ref{DLTF}))
\beq\label{DLTFlam}
\Delta (t\gg 1/\omega_p)\,\simeq\, {\rm exp}\left\{ -\alpha Tt\Bigl(
 \ln(\omega_p t)\,+\,0.12652...
\,+\,\lambda/2\Bigr)\right\}.\eeq
(The gauge-dependent piece in the exponent coincides
with the corresponding piece of the one-loop
damping rate, $\gamma_{\lambda}\,=\,
{\lambda}\,\alpha T/2$  \cite{Schiff92,Rebhan93}.)

However, if we consider the large-time
behaviour at fixed $\mu$, then we can use the asymptotic
expansion of ${\rm si}(x)$, that is
\beq
-\,{\rm si}(x\gg 1)\,\sim\,\frac{1}{x}\,\left(\cos x\,+\,\frac
{\sin x}{x}\,+\,{\cal O}(1/x^2)\right),\eeq
to obtain
\beq\label{Flam22}\
F_\lambda(t \gg 1/\mu)\,=\,\frac{1}{8\pi^2 \mu} \,\left[
3\,-\,\frac {\sin 2\mu t}{2\mu t}\,+\,{\cal O}(1/\mu^2 t^2)\right].\eeq
In this case, the sole effect of the gauge-dependent piece $F_\lambda(t)$
at times $t \gg 1/\mu$ is to change the normalization
of the propagator, by a factor
\beq\label{resid}
{\rm exp}\Bigl(-\lambda g^2 T F_\lambda(t)\Bigr)\,\simeq\,
{\rm exp}\left(-\lambda\, \frac{3 \alpha}{2\pi}\,
\frac{T}{\mu}\right)\,\equiv \,z(T,\mu,\lambda),\eeq
which is both gauge-dependent and cutoff-dependent.

The gauge-dependent contribution to the damping rate,
eq.~(\ref{Flam11}), arises because the on-shell fermion 
is kinematically allowed to ``decay''
with the emission, or the absorption, of a massless gauge ``photon''.
At $T=0$, such an emission process cannot occur:
by kinematics, the emitted photon must be colinear
($\theta = 0$, $q_0=q$), 
and the corresponding phase space vanishes. But this is not so
at finite temperature, because of
the Bose-Einstein factor $N(q_0)\sim T/q_0$ which diverges
as $q_0\to 0$ (see Appendix B of Ref. \cite{prd})
for an explicit calculation). After HTL resummation, the gauge sector
is the only one to contain massless fields.  The unphysical decay channel
can be suppressed by giving the gauge photon a small mass $\mu$,
as originally proposed by Rebhan \cite{Rebhan93}
(see also Refs. \cite{Sen,debye}). As we have seen, this procedure
 ensures the gauge-independence of the  damping rate, to the
order of interest.

Further insight may be gained by a comparison with
the corresponding results at zero temperature \cite{Bogoliubov}. 
After ultraviolet renormalization, the retarded BN propagator 
at zero temperature is given by 
\beq\label{SRT0}
S(t,{\bf p})\,\propto \, {\rm e}^{-iE_pt}\,
{\rm exp}\left\{(3-\lambda)\,\frac{\alpha}{2\pi}\,\ln(mt)\right \}\,=\,
\left(mt\right)^{(3-\lambda)\,\frac{\alpha}{2\pi}}
{\rm e}^{-iE_pt}\,.\eeq
Thus, in the energy representation, 
 the mass-shell singularity is generally a branch point,
rather than a simple pole:
\beq\label{SRT0P}
S(p)\,\propto \,\frac{1}{u\cdot p - m}
\left(\frac{m}{u\cdot p - m}\right)^{(3-\lambda)\,\frac{\alpha}{2\pi}}\,,\eeq
where $u^\mu=(u_0,{\bf u})$ is the fermion four-velocity,
$u^\mu =p^\mu/m$, with $u^2=1$. 

Note that both the physical and the gauge sectors
of the photon propagator contribute to the
mass-shell behaviour in eq.~(\ref{SRT0P}):
 at $T=0$, the gauge field quanta
are massless in both sectors. 
Furthermore, no infrared regulator is necessary: 
in deriving eqs.~(\ref{SRT0})--(\ref{SRT0P}), one encounters
no IR divergences, and the position of the mass shell is 
gauge-independent, as it should. Still, if one wishes to perform
soft-photon computations in perturbation theory in any other gauge
than the Yennie gauge ($\lambda=3$), it is convenient
to introduce an infrared regulator, so as to recover
the simple-pole structure of the mass-shell.
When the photon is given a small mass $\mu$,
 the propagator (\ref{SRT0P}) is replaced by
\beq\label{SRT1P}
S(p)\,\propto \,\frac{z(\mu,\lambda)}{u\cdot p - m}\,\eeq
where the residue
\beq z(\mu,\lambda)\,=\,
{\rm exp}\left\{(3-\lambda)\,\frac{\alpha}{2\pi}\,\ln(m/\mu)\right\}\,,\eeq
is gauge-fixing dependent and also cutoff-dependent. It may be compared
to the finite-temperature normalization factor in eq.~(\ref{resid}). 

We see that, as a consequence of the Bose-Einstein enhancement of the soft
photon processes, the divergence of the ``residue'' $z(T,\mu,\lambda)$ 
as $\mu \to 0$ is linear at $T>0$,  rather than just logarithmic
at $T=0$. Moreover, if at $T=0$ the introduction of a photon mass
is just a matter of convenience, at $T>0$
the use of an  infrared regulator in the gauge sector is compulsory
in order to eliminate the contribution of the non-physical degrees
of freedom and avoid the gauge-dependence of the mass-shell.

\section{Some results for QCD}
\setcounter{equation}{0}

We consider now the generalization of the previous arguments
to the non-Abelian case, that is, to the high-temperature,
weakly-coupled ($g(T)\ll 1$) quark-gluon plasma.
The self-interactions of the  soft gluons prevent us from
getting in this case an explicit solution.
However, it is expected \cite{Linde80} that these interactions
generate screening of the static magnetic fields.
If this is so, the corresponding screening length,
typically of order $1/g^2T$, provides then a natural IR cutoff
of order $g^2 T$ in the perturbation theory for $\gamma$.

We shall investigate this possibility in the next
subsection, in the framework of a toy model which is QED with an
infrared cutoff $\mu \sim g^2T$ in the magnetic sector.
By solving this model in the BN approximation,
we shall obtain a qualitative picture of the effects of
the magnetic mass on the large-time behaviour of the fermion
propagator.

Then, we shall propose a functional integral representation
for the propagator of a hard quark or gluon which, being
formulated in three-dimensional Euclidean space, is a priori
well suited for lattice calculations. This formulation allows
for a direct numerical study of the particle decay in real time.

\subsection{QED with a magnetic mass}

To implement magnetic screening in QED, we replace
the massless static transverse propagator by its massive version,
\beq\label{MM}
D_{ij}(0,{\bf q})\,=\,\frac{\delta_{ij}-\hat q_i\hat q_j}
{q^2+\mu^2}\,,\eeq
with $\mu \sim g^2T$. Of course, such an infrared behaviour
could not occur in QED, where the correct magnetic polarization
tensor $\Pi_t(0,q)$ vanishes like $q^2$ when $q\to 0$, to all orders
in perturbation theory \cite{Fradkin65}. We simply use
``massive QED'', as defined by eq.~(\ref{MM}),
as a crude parametrization of the non-perturbative
screening effects in QCD. 

Strictly speaking, when $\mu >0$ we have no infrared
divergences. However, as long as $\mu\simle g^2T$, the dominant
contribution to the damping rate is still given by the static
magnetic photons. For instance, to one loop order
the static mode yields
\beq\label{logM}
\gamma_{st}\,=\,\alpha T\,\ln\frac{\omega_p}{\mu}\,\sim\,g^2 T
\ln(1/g),\eeq
which is enhanced by a factor $\ln(1/g)$ as compared to
the  contribution of the non-static modes (which
is $\sim g^2T$, as in eq.~(\ref{DLTF})). Moreover,
the higher-loop diagrams 
contribute terms of relative order $(\alpha T/\mu)^{n-1}$,
where $n$ is the number of loops, so that the perturbation
theory breaks down for  $\mu\simle g^2T$ \cite{prd}.
This is why a non-perturbative calculation is necessary even in
the presence of an infrared cutoff  $\mu\simle g^2T$.

In order to get the leading contribution
to the damping factor, we can restrict ourselves
to the interactions with {\it static} ($q_0=0$) magnetic photons
\cite{prl,prd}.
In practice, such a calculation amounts to preserve only the
contribution $F_{IR}(t)$ in eq.~(\ref{FIR}),
where however the massive propagator (\ref{MM}) must now be used.
This gives:
\beq\label{DMU}
\Delta_\mu(t)\,\simeq\,{\rm exp}\left\{-g^2T
\int \frac{{\rm d}^3q}{(2\pi)^3}
\,\frac{1}{q^2+\mu^2}\,\frac{1-  {\rm cos}\,t({\vq})}{({\vq})^2}
\right\}\,.\eeq
As explained in section 3,
the integral in  eq.~(\ref{DMU}) has to be computed
with an upper cutoff $\omega_p \sim  gT$, to account approximately
for the effect of the neglected non-static modes.
(In  eq.~(\ref{FIR}), the upper cutoff was provided by
the substracted term $1/(q^2+\omega_p^2)$. In eq.~(\ref{qint} below,
we shall find convenient to introduce this cutoff in a different way.
At large times $\omega_p t \gg 1$, the leading contribution
to the damping factor is indeed insensitive to
the precise value of the UV cutoff, and also to the specific procedure
which is used for its implementation \cite{prd}.)

\begin{figure}
\protect \epsfxsize=12.cm{\centerline{\epsfbox{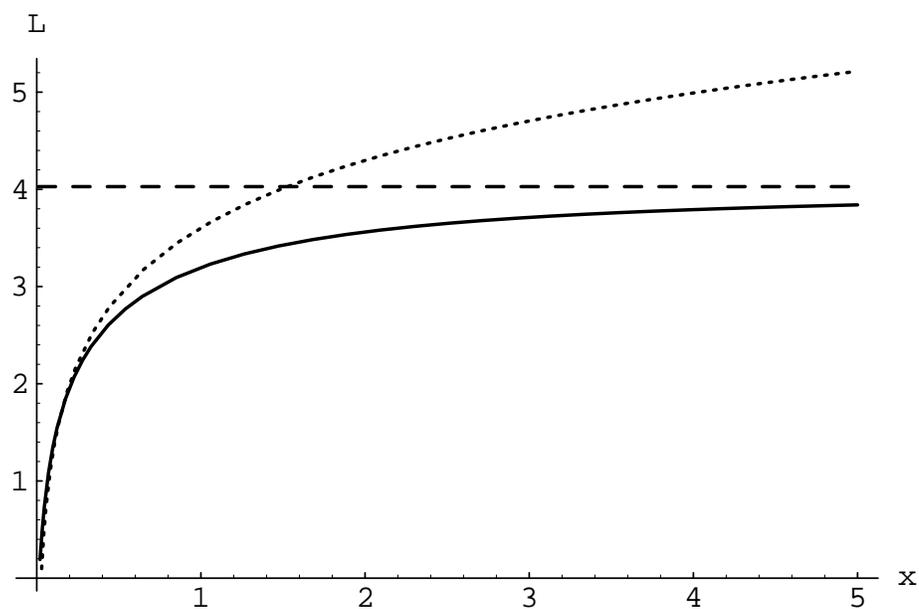}}}
	 \caption{The function $L(x,y)$, eq.~(\ref{Lxy}),
is represented as a function of $x$ for fixed $y=0.01$ (continuous
line). We have also represented the asymptotics $L_1(x,y)\equiv
\ln(x/y) - 1$ (dotted line) and  $L_2(x,y)\equiv \ln(1/y) - \gamma_E$
(dashed line). These are good approximations to $L(x,y)$ in the
domains $y\ll x\ll 1$ and $x\gg 1$, respectively.}
\label{Lmu}
\end{figure}

To perform the integral in eq.~(\ref{DMU}), we write
$\Delta_\mu(t)={\rm exp}(-g^2T F_\mu(t))$, with
\beq\label{qint}
F_\mu(t)&=&
\frac{1}{2}\int_0^t {\rm d}s_1 
\int_0^t {\rm d}s_2\,
\int \frac{{\rm d}^3q}{(2\pi)^3}
\,\frac{{\rm e}^{i({\bf v\cdot q})(s_1-s_2)}}{
q^2+\mu^2}\,\nonumber\\&=&
\frac{1}{8\pi}\int_0^t {\rm d}s_1 
\int_0^t {\rm d}s_2\,\, \frac{
{\rm e}^{-\mu|s_1-s_2|}}{|s_1-s_2|}\,
\,\theta(|s_1-s_2|-1/\omega_p)\nonumber\\&=&
\frac{1}{4\pi}\int_{1/\omega_p}^t
 {\rm d}s\,\frac{t-s}{s}\,{\rm e}^{-\mu s}\nonumber\\&=&
\frac{t}{4\pi}\left \{
\int_{\mu/\omega_p}^{\mu t} \frac{{\rm d}x}{x}\,{\rm e}^{-
x}\,-\,\frac{{\rm e}^{-\mu/\omega_p}-
{\rm e}^{-\mu t}}{\mu t}\,\right \}.\eeq
In this calculation, the  ultraviolet cutoff
has been introduced, in the second line, in the function
 $\theta(|s_1-s_2|-1/\omega_p)$. For the purpose of a graphical 
representation (see  Fig.~\ref{Lmu}), we rewrite the final expression
 above as $F_\mu(t)\equiv (t/4\pi) L(x,y)$,
with $x\equiv \mu t$, $y\equiv \mu/\omega_p$, and
\beq\label{Lxy}
L(x,y)\,\equiv \,{\rm E}_1(y)\,-\,{\rm E}_1(x)
\,-\,\frac{{\rm e}^{-y}-{\rm e}^{-x}}{x}\,,\eeq
where ${\rm E}_1(x) $ is the exponential-integral function \cite{Ober},
 ${\rm E}_1(x)=\int_1^\infty {\rm d}z \,({\rm e}^{-xz}/z)$.
For  $\mu\sim g^2T$, we have $y\sim g \ll 1$.
We recall that the above calculation  only makes sense for
large enough times, $\omega_p t\gg 1$ or $x \gg y$.

Since the expression in eq.~(\ref{qint}) involves two energy scales, 
namely $\omega_p$ and $\mu$, with $\mu\ll\omega_p$, we distinguish
between two regimes of time: (i)  very large times, $t\gg 1/\mu$
(i.e., $x\gg 1$), where
\beq\label{qLT1}
F_\mu(t)\simeq \,\frac{t}{4\pi}\,\Bigl(\ln \frac{\omega_p}{\mu}
\,+\,{\cal O}(1)\Bigr),\eeq
and (ii) intermediate times,  $1/\omega_p \ll t \ll 1/\mu$
(i.e., $y\ll x\ll 1$), where
\beq\label{qIT}
F_\mu(t)\simeq \,\frac{t}{4\pi}\,\Bigl(\ln({\omega_p}t)
\,+\,{\cal O}(1)\Bigr).\eeq
Thus, the ``magnetic mass'' $\mu$ is only felt at sufficiently
large times --- where $\Delta_\mu(t)$ decays exponentially
in agreement with the one-loop result (\ref{logM}) ---
while it has no effect at intermediate times.
We remark at this point that, when discussing the lifetime of the 
excitation, it is rather the intermediate times which matter, since for
asymptotically large times $t\simge 1/g^2T \sim 1/\mu$ the excitation
has already decayed. 
This behaviour, eqs.~(\ref{qLT1})--(\ref{qIT}), can be also observed
in Fig.~\ref{Lmu}, where we have represented $L(x,y)$ as
a function of $x$ for a fixed, small, value of $y$ (namely $y=0.01$).
For $x$ of order one, one clearly
 sees on this figure the transition between the two types of behaviour,
as described by eq.~(\ref{qLT1}) and eq.~(\ref{qIT}), respectively.

It has been suggested, first by Lebedev and Smilga
\cite{Lebedev90}, that when computing the damping rate to one-loop order,
the damping rate itself should be self-consistently
resummed in the internal hard line. The usual  argument goes as follows:
since $\gamma \sim g^2T$ is of the same order as the infrared cutoff  $\mu$,
it should be taken into account when studying the infrared behaviour
of the integrand. If we do that, by following Ref. \cite{Pisarski93},
 then the one-loop result (\ref{logM}) is modified to
\beq\label{logM1}
\gamma\,\simeq\,\frac{\alpha T}{2}\,\ln\frac{\omega_p^2}{\mu^2+
2\mu\gamma}\,.\eeq
(Up to appropriate color factors, the same result is obtained in 
QCD, for both quarks and gluons  \cite{Pisarski93}.)
However, this is not correct:
the self-energy resummation advocated in the procedure leading
to eq.~(\ref{logM1}) should be accompanied by a corresponding
resummation in the vertex function, so as to respect gauge symmetry.
As discussed in Ref. \cite{prd}, the vertex corrections
 generate new infrared divergences, and, when added to
the self-energy corrections, 
conspire to give a leading-order estimate for the damping rate
which has precisely the form indicated in  eq.~(\ref{logM}).
(See Appendix C in Ref. \cite{prd} for more details.) At this point,
it might be useful to emphasize that the BN calculation provides
precisely a self-consistent resummation of the fermion propagator
near the mass-shell, together with the appropriate resummation of 
the vertex function, as required by gauge symmetry.

\subsection{QCD}

Going now to QCD, we first observe that
the Bloch-Nordsieck approximation remains relevant
 to discuss the large-time (or mass-shell) behaviour
of the quasiparticle propagator, and this for both
quarks and (transverse) gluons. (We consider here a hard
quasiparticle, with momentum $p\simge T$.)
While for quarks this approximation is  
easy to justify, by analogy to QED, the case of gluons
requires more care and is discussed in Appendix C.
Moreover, we expect the leading large-time behaviour to be given by
the quasiparticle interactions with {\it static} ($q_0=0$) and
very soft ($q\to 0$) magnetic gluons:
indeed, these are the interactions which generate the infrared
divergences of the perturbation theory \cite{prd}.

What is new with respect to QED, is that the relevant self-energy
corrections also include the mutual interactions of the internal
gluons, expected, in particular, to lead to magnetic screening.
\begin{figure}
\protect \epsfxsize=16.cm{\centerline{\epsfbox{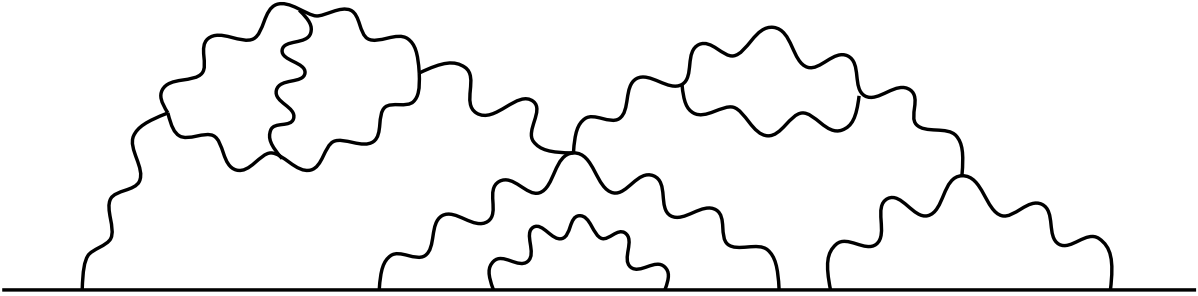}}}
	 \caption{A generic self-energy diagram in QCD which yields
infrared divergences on the mass-shell. The continuous line
is a hard particle (quark or gluon). The wavy lines are soft
magnetostatic gluons. All the loop integrations are three-dimensional.}
\label{QCD}
\end{figure}
A typical Feynman graph contribution to the self-energy
is depicted in Fig.~\ref{QCD}. 
The continous line in this diagram is hard, and may represent
either a quark, or a gluon. The wavy lines denote static magnetic
gluons, and all the loop integrations are three-dimensional.
As explained in section 3, these integrations involve
an upper cutoff $\omega_p$. In QCD \cite{BIO96},
\beq
\omega_p^2\,=\,\frac{g^2 T^2}{18}\,(2N+N_{\rm f}),\eeq
for $N$ colours and $N_{\rm f}$ flavours of thermal quarks.
(Note that the gluonic loops attached to soft internal lines
would be finite even in the absence of the ultraviolet cutoff.)

In the BN approximation, all such diagrams are formally
 resummed by the following functional integral 
(the gauge-fixing terms are not written here explicitly;
see the discussion below, after eq.~(\ref{Delta0N})):
\beq\label{FS}
S(x-y)= Z^{-1}\int {\cal D} {\bf A}\, G(x,y|{\bf A})\,{\rm
exp}\left\{-\frac{1}{4T}\int {\rm d}^3x F_{ij}^a F_{ij}^a\right\},\eeq
 where ${\bf A}^a({\bf x})$ is a static color field,
$F^a_{ij}=\del_i A_j^a - \del_j A_i^a - g f^{abc} A^b_i A^c_j$,
($f^{abc}$ are the structure constants of the colour group),
 and $ G(x,y|{\bf A})$ satisfies the equation
\beq\label{SAN}
i (v\cdot D_x)  G(x,y|{\bf A})=\delta^{(4)} (x-y),\eeq
where $D_\mu=\del_\mu +igA_\mu$, $A_\mu=(0,{\bf A})$, and
$A_i\equiv A^a_i T^a$ is a color matrix in either the adjoint
or the fundamental representation (for gluons or quarks, respectively).
The plasma effects do not modify the gluonic action in eq.~(\ref{FS})
(recall that the HTL-corrections vanish
for static magnetic fields \cite{BP90}), but only enter
 through the upper cutoff $\omega_p\sim gT$.

The solution of the BN equation (\ref{SAN})
with retarded boundary conditions is immediate \cite{BIO96}:
\beq\label{GRN}
G_R(x,y|{\bf A})&=&-i\,\theta (x_0-y_0)\,\delta^{(3)}
\left({{\bf x}}-{{\bf y}}-{{\bf v}}(x_0-y_0)
\right )U({\bf x, y})\nonumber\\
U({\bf x,x-v}t)&=&{\rm P}\,\exp\left\{ ig\,\int_0^t {\rm d}s\, {\bf v}
\cdot {\bf A}({\bf x-v}(t-s))\right\},\eeq
where  the path-ordering operator P is necessary since the color
 matrices ${\bf A}({\bf x})$ at different points along the
path do not commute with each other.

The retarded propagator $S_R(x-y)$ is calculated by inserting 
eq.~(\ref{GRN}) in  the functional integral (\ref{FS}). 
It can be written as 
\beq\label{SRTN}
S_R^{ab}(t,{\bf p})&=&-i\,\delta^{ab}
\theta(t)\, {\rm e}^{-it({\vp})}\,\Delta(t),\eeq
with
\beq\label{Delta0N}
\Delta(t)\equiv {\cal N}
\int {\cal D} {\bf A}\,{\rm Tr} \,U({\bf x,x-v}t)\,{\rm
exp}\left\{-\frac{1}{2T}\int {\rm d}^3x \,{\bf B}^2 \right\},\eeq
${\bf B}^2=B^a_i B^a_i$, and $B^a_i=(1/2)\epsilon_{ijk} F^a_{jk}$ is 
the chromomagnetic field. In order for eq.~(\ref{Delta0N}) to be
well-defined, it is further necessary to choose a gauge within
the functional integral.
(Recall that the parallel transporter
${\rm Tr} \,U({\bf x,x-v}t)$ is not invariant under the gauge
transformations of the background color field.)
However, we shall argue below that the dominant large-time
behaviour should be independent of the gauge-fixing condition.

The functional integral (\ref{Delta0N}) cannot be computed
analytically because ${\bf B}^a$ is non-linear in the gauge potentials.
However,  we may expect the large-time behaviour
of $\Delta(t)$ to be similar to that of the model
discussed in section 5.1, that is (cf. eqs.~(\ref{qLT1})--(\ref{qIT})),
\beq\label{qIT2}
\Delta(1/\omega_p \ll t\ll 1/\mu)\simeq {\rm exp}\left\{
- \,C_r\,\frac{g^2T}{4\pi}\,t\,\Bigl(\ln(\omega_p t)
\,+\,{\cal O}(1)\Bigr)\right\},\eeq
at intermediate times, and, respectively,
\beq\label{qLT2}
\Delta(t\gg 1/\mu)\simeq {\rm exp}\left\{
- \,C_r\,\frac{g^2T}{4\pi}\,t\,\Bigl(\ln \frac{\omega_p}{\mu}
\,+\,{\cal O}(1)\Bigr)\right\},\eeq
at very large times. In these equations, $C_r$ is the Casimir
factor of the appropriate color representation
(i.e.,   $C_q =(N^2-1)/2N$ for a hard quark, and $C_g=N$ for 
a hard gluon), and the magnetic mass $\mu\sim g^2T$ is expected
to come out from the soft gluon mutual interactions.
Note that the ${\cal O}(1)$ terms in the above equations
are not consistently determined by the present approximation:
Indeed, from the experience with QED, and also
from the one-loop calculations in QCD \cite{Lebedev90,Pisarski93,all},
we know that such terms receive contributions
from the non-static gluon modes, and that they may
 be sensitive to the gauge-fixing condition (cf. section 4).

To verify this picture, one could rely on a lattice computation
of the Euclidean functional integral (\ref{Delta0N}). 
The parallel transporter ${\rm Tr} \,U({\bf x,x-v}t)$ is easily
implemented as a product of link operators.
(Recall that ${\bf v}$ is a fixed unit vector,
e.g., ${\bf v}=(0,0,1)$, so that $U({\bf x,x-v}t)$ is
a product of link operators in the $z$ direction, from
$z_0-t$ to $z_0$, with $z_0$ an arbitrary site on the lattice.)
Since the expression (\ref{Delta0N}) is defined with an upper cutoff
$\omega_p\sim gT$, the lattice spacing $a$ is fixed: $a\sim 1/\omega_p$.
From perturbation theory,  we expect 
the decay of $\Delta(t)$ at times $t\gg a$ to be only logarithmically sensitive
to the precise value of $a$ (cf.  eqs.~(\ref{qIT2})--(\ref{qLT2})).

The objective of a lattice calculation would be then to verify
the large-time behaviour predicted in eqs.~(\ref{qIT2})--(\ref{qLT2})).
By observing the interplay between these two regimes, one may verify what
is the typical scale for the emergence of magnetic screenig: indeed,
we expect the transition between two regimes to occur for $t \sim 1/\mu$.
However, this could not be sufficient for a quantitative measure
of the magnetic mass, because of the theoretical uncertainty
on the subleading term ${\cal O}(1)$.

The main limitation against an explicit calculation comes from the
lattice size: indeed, in order to verify the aforementioned picture,
 one needs a small coupling constant
$g\ll 1$ --- to ensure a clean separation between
the scales $g^2T$ and $gT$ --- together with large values of time,
up to $t_{max}\gg  1/g^2 T \sim a/g$.
Thus, the lattice should have at least $N^3$ sites,
with $N\simge t_{max}/a \gg 1/g$.

\section{Conclusions}
\setcounter{equation}{0}

In this paper, we have completed the analysis of the 
Bloch-Nordsieck propagator in hot QED, and we have
also discussed the usefulness of such an approximation for
a high-temperature QCD plasma.

As compared to Refs. \cite{prl,prd}, several
points have been clarified by the present analysis.
First, the three-dimensional model of \cite{prl,prd}
suffers from a spurious ultraviolet divergence coming
from the restriction to the static photon mode. We have shown here
 that the contribution of the non-static modes
provides a dynamical cutoff at momenta $\sim gT$. Not only this justifies
the cut-off procedure used in Refs. \cite{prl,prd},
but it also allows one to compute explicitly the subleading term
in the large time behaviour (cf. eq.~(\ref{DLTF})).

Secondly, the effects of the gauge-fixing procedure
 only enter at the level of the subleading term.
Thus, by computing this term in different gauges, one
can study the gauge-(in)dependence of the large-time decay.
We have performed this computation in the Coulomb gauge,
and in a generic covariant gauge, with conclusions
which agree with the one-loop calculations in
Refs. \cite{Schiff92,Rebhan93}; namely, the subleading term
is gauge-independent if computed in the
presence of an infrared cutoff in the gauge sector.
Physically, such a cutoff  separates the particle
mass-shell from the threshold for the spurious
emission or absorbtion of massless gauge photons.
With a non-zero infrared regulator $\mu$, the gauge-dependent
contribution to the damping vanishes in a vicinity $\sim\mu$
 of the mass-shell, or, equivalently, for times
$t\gg 1/\mu$. Thus, to avoid spurious gauge contributions
 over a particle lifetime, one should choose $\mu\simge \gamma$, where
 $\gamma \sim g^2T\ln(1/g)$.

Concerning QCD, we have argued that the BN model may still be a relevant
approximation for the study of the quasiparticle mass-shell.
As compared to QED, the non-Abelian model is complicated by the soft
gluon self-interactions.
The problem simplifies considerably when one considers only
the dominant contribution due to the static magnetic gluon
 modes: then, not only all the HTL 
corrections vanish, but the needed path-integral can in
principle be computed on a three-dimensional lattice,
 with a fixed lattice spacing $a\sim 1/gT$.

\vspace*{1.cm}
{\noindent {\large{\bf Acknowledgements}}}

It is a pleasure to thank S. Belyaev, A. Krasnitz,
L. McLerran, J.-Y. Ollitrault and B. Vanderheyden
for discussions and useful remarks on the manuscript.

\bigskip
\bigskip
\bigskip

\setcounter{equation}{0}
\renewcommand{\theequation}{A.\arabic{equation}}
\appendix{\noindent {\large{\bf Appendix A}}}

In this appendix, we construct the BN propagator
for a  thermal fermion with momentum $p\sim T$. The final result
turns out to be essentially the same as that obtained in 
section 3 in the case $p\gg T$.

We use the imaginary-time formalism
which has been developed in Ref. \cite{prd}. In this formalism,
we have to solve the imaginary-time BN equation (cf. eq.~(\ref{SA})):
\beq\label{SAtau}
- (v\cdot D_x)  G(x,y|A)=\delta_E (x,y),\eeq
with antiperiodic boundary conditions (cf. eq.~(\ref{BC})):
\beq\label{BCG} G_E(\tau_x=0,\tau_y|A)=- G_E(\tau_x=\beta,\tau_y|A),\eeq
and similarly for $\tau_y$. In these equations, the time
variables are purely imaginary ($x_0=t_0-i\tau_x$ and $y_0=t_0-i\tau_y$,
with $0\le \tau_x,\, \tau_y \le \beta$ and
$\delta_E(x-y)= \delta(\tau_x-\tau_y)\delta({\bf x-y})$),
and the gauge fields are periodic in imaginary time:
 $A_\mu(\tau=0)=A_\mu(\tau=\beta)$.
In Ref. \cite{prd}, we have solved this equation explicitly
in perturbation theory, i.e., as a series in powers of $gA$,
and then we have performed the functional integration
over the gauge fields (cf. eq.~(\ref{FunctionalS})).
The resulting propagator can be written as (cf. eq.~(\ref{SC1}))
\beq\label{SE0}
iS(x_0-y_0,{\bf p})&=&\theta(\tau)S^>(\tau, {\bf p})-
\theta(-\tau)S^<(\tau, {\bf p}),\eeq
where the analytic functions $S^>$ and $S^<$ are obtained 
in the form \cite{prd}
\beq \label{S<}
S^<(\tau, {\bf p})&=& {\rm e}^{-\tau E_p} \tilde V(E_p; u=-\tau)
\qquad {\rm for}\,\,\,\,-\beta \le \tau\le 0,\nonumber\\
S^>(\tau, {\bf p})&=&
 {\rm e}^{(\beta-\tau) E_p } \tilde V(E_p; u=\beta-\tau)
\qquad {\rm for}\,\,\,\,0\le \tau\le \beta.\eeq
In this equation, $E_p\equiv {\bf v\cdot p}$ is the BN mass-shell
and the function $\tilde V(E_p;u)$ is given, for $
0\le u \le \beta$, as a formal series in powers of $g^2$ \cite{prd}:
\beq\label{tilVpert}
\tilde V(E_p;u)&=&n(\vp)+
\sum_{n\ge 1}(-1)^n\frac{g^{2n}}{n!}
\int[{\rm d}q_1 {\rm d}q_2\,...\,{\rm d}q_n]
\frac{\tilde D(q_1) \tilde D(q_2)\,...\,\tilde D(q_n)}
{(v\cdot q_1)^2 (v\cdot q_2)^2\,...\,(v\cdot q_n)^2}\nonumber\\
&{}&\Bigl[n(\vp)-n({\bf v\cdot (p +q_1)})\,
{\rm e}^{-u(v\cdot q_1)}-n({\bf v\cdot (p+q_2)})\,{\rm e}^{-u(v\cdot q_2)}
+\nonumber\\ &{}&\,...\,+(-1)^n\, n({\bf v\cdot (p +q_1+q_2+...+q_n)})\,
{\rm e}^{-u\,v\cdot(q_1+ q_2+...+ q_n)}\Bigr]
\,,\eeq
with $\tilde D(q)\equiv v^\mu\, {}^*D_{\mu\nu}(i\omega_m,{\bf q}) v^\nu\,$.
In this and the following  equations, the photon energies $q_i^0$
are discrete and purely imaginary: $q^0=i\omega_m=i 2\pi mT$,
with integer $m$ (Matsubara frequencies).
The measure in the momentum integrals is denoted by
\beq\int[{\rm d}q]\equiv T\sum_{\omega_m}
\int\frac{{\rm d}^3q}{(2\pi)^3}\,.\eeq

The thermal factors make the momentum integrals
in eq.~(\ref{tilVpert}), like
\beq
\int\frac{{\rm d}^3q}{(2\pi)^3}\,n({\vpq})\,{\rm e}^{u({\bf v\cdot q})},\eeq
 convergent for any $0< u <\beta$.
This, in turn, ensures the analyticity of
the functions $S^<(\tau)$ and  $S^>(\tau)$ in (\ref{S<}) \cite{MLB96}. By
 analytically continuing these functions toward
the real-time axis (i.e., by replacing $\tau\to it$ with real $t$),
one constructs the retarded propagator:
\beq\label{SR}
S_R(t,{\bf p})&=&-i\theta(t)\Bigl(S^>(t, {\bf p})+S^<(t, {\bf p})\Bigr)\nonumber\\
&=&-i\theta(t) \,{\rm e}^{-itE_p}\Bigl\{
  {\rm e}^{\beta E_p} \tilde V(E_p; u=\beta-it)
+ \tilde V(E_p; u=-it) \Bigr\} .\,\eeq
Note, however, that the analytic continuation to real time
can be done only after performing the Matsubara sums in all the
terms of the series in eq.~(\ref{tilVpert}).

Fortunately, this can be done easily in the relevant regime
of large-time ($t\gg 1/gT$). According to the discussion in section 3,
we expect then the momentum integrals to be dominated by {\it soft}
photon momenta, $q\simle gT$. Indeed,
the photon propagator $\tilde D(q)$, which can be rewritten as
(with $ \tilde \rho(q)\equiv v^\mu\rho_{\mu\nu}(q)v^\nu$;
cf. eq.~(\ref{Coul})) :
\beq\label{Sspec}\tilde D
(i\omega_m, {\bf q})&=&\int_{-\infty}^{\infty}\frac{{\rm d}\omega}{2\pi}
\,\frac{\tilde\rho(\omega, q)}{\omega-i\omega_m}\,,\eeq
provides, through the spectral density $\tilde\rho(\omega,{\bf q})$,
an effective upper cutoff  $\sim gT$ for the integrals over ${\bf q}$.
(Recall that the functions $\beta_l(\omega,q)$
and $\beta_t(\omega,q)$ in eq.~(\ref{RRHO})
 are rapidly decreasing for $q\gg gT$.)
Strictly speaking, this cutoff becomes effective only {\it after}
$u$ is continued to $\beta-it$ or $-it$. However, we may anticipate
for its effect and supply the integrals
over ${\bf q}_i$ in eq.~(\ref{tilVpert}) with an upper cutoff
$\sim gT$. Then the photon momenta are limited to values
 $|{\bf q}|\ll |{\bf p}|\sim T$, and we can replace
$ n({\bf v\cdot (p+q)})$ by $n({\bf v\cdot p})$ up to terms
of order $q/T\simle g$. The fermion occupation factor
 $n({\bf v\cdot p})$ in
eq.~(\ref{tilVpert}) then factorizes, and the resulting 
expression can be resummed into an exponential:
\beq\label{tilV}
\tilde V(E_p;u)&\approx &n(E_p)\,\Delta^<(u)\nonumber\\
\Delta^<(u)&\equiv& {\rm exp}\left\{-g^2 \int[{\rm d}q]\,\tilde D(q)\,
\frac{1\,-\,{\rm e}^{-u(v\cdot q)}}{(v\cdot q)^2}\right\}.\eeq
At this stage, we can then perform  the Matsubara sum over $q^0=i\omega_m$ 
(by using the spectral representation in eq.~(\ref{Sspec}),
together with contour integration), and obtain
\beq\label{DEu}
\Delta^<(u)\,=\, {\rm exp}\left\{-g^2 \int \frac{{\rm d}^4q}{(2\pi)^4}
\,\tilde\rho(q_0,{\bf q})\left[(1+N(q_0))\,
\frac{1-{\rm e}^{-u(v\cdot q)}}{(v\cdot q)^2}\,-\,
(1+N({\vq}))\,\frac{u}{v\cdot q}\right]\right\},\nonumber\\\eeq
where $v\cdot q = q_0-{\vq}$ (we have renamed $q_0$  the real energy $\omega$).
The last expression can be now continued to $u\to -it$,
with the result:
\beq\label{DeltaE}
\Delta^<(-it)&\equiv& {\rm exp}\left\{-g^2 \int \frac{{\rm d}^4q}{(2\pi)^4}
\,\tilde\rho(q)\left[(1+N(q_0))\,
\frac{1-{\rm e}^{it(v\cdot q)}}{(v\cdot q)^2}\,+\,
(1+N({\vq}))\,\frac{it}{v\cdot q}\right]\right\}\nonumber\\
&=& {\rm exp}\{-it\Phi(t)\}\,\Delta(t),\eeq
which involves the same functions $\Phi(t)$ and $\Delta(t)$ 
as in section 3 (cf. eq.~(\ref{PHI}) and (\ref{Delta2})).
At this point, the momentum integral in eq.~(\ref{DeltaE})
is ultraviolet finite
and the cutoff can be removed. Recalling eq.~(\ref{S<}), we can
 finally write:
\beq\label{S<f}
S^<(t, {\bf p})\,=\,{\rm e}^{-itE_p} n(E_p)\,\Delta(t),\eeq
where we have ignored the inconsistent phase $\Phi(t)$.

To compute $S^>(t, {\bf p})$, we start with (cf. the second
eq.~(\ref{S<})):
\beq
 {\rm e}^{\beta E_p } \tilde V(E_p; u=\beta-\tau)
\,\approx \,[1-n(E_p)]\,\Delta^>(\tau)\nonumber\\
\Delta^>(\tau)\,\equiv\,{\rm exp}\left\{-g^2 \int[{\rm d}q]\,\tilde D(q)\,
\frac{1\,-\,{\rm e}^{\tau (v\cdot q)}}{(v\cdot q)^2}\right\},\eeq
where $1-n(E_p)\equiv {\rm e}^{\beta E_p}n(E_p)$ has been
factorized by the same approximations as above.
After performing  the Matsubara sum and the analytic continuation
$\tau\to it$, we finally obtain (within the same accuracy as in 
eq.~(\ref{S<f})):
\beq\label{S>f}
S^>(t, {\bf p})\,=\,{\rm e}^{-itE_p}[1- n(E_p)]\,\Delta(t).\eeq
Thus, for sufficiently large times, both functions $S^<(t)$ and $S^>(t)$
decay like $\Delta(t)$, eq.~(\ref{Delta2}). The same is therefore
true for the retarded propagator, as given by eqs.~(\ref{SR}),
(\ref{S<f}) and (\ref{S>f}):
\beq\label{SR12}
|S_R(t,{\bf p})|&\propto&\Delta(t),\eeq
which is the result quoted in eq.~(\ref{SR1}).

\newpage
\setcounter{equation}{0}
\renewcommand{\theequation}{B.\arabic{equation}}
\appendix{\noindent {\large{\bf Appendix B}}}

In this appendix, we calculate the double integral
in eq.~(\ref{FregLT}), thus proving the result quoted in eq.~(\ref{FregF}).
The method to be used here was suggested to us by Jean-Yves Ollitrault
(see also \cite{JYB}). We first write
\beq\label{I}
I\,\equiv\, \int_0^\infty {\rm d}q \,q
 \int_{-q}^q \frac{{\rm d}q_0}{2\pi q_0}\left(
\beta_l(q_0,q) \,-\,\frac{q_0^2}{q^2}\,\beta_t(q_0,q)
+ \nu_t(q_0,q)\right)\,\equiv\,I_1+I_2+I_3,\eeq
where the three pieces $I_s$, $s=1,\,2,\,3$, correspond to the
three terms within the integrand.
To illustrate the method, we compute the second piece in detail:
\beq\label{I20}
I_2&=&- \int_0^\infty {\rm d}q \,q
 \int_{-q}^q \frac{{\rm d}q_0}{2\pi q_0}\,\frac{q_0^2}{q^2}\,\beta_t(q_0,q)
\nonumber\\&=&-
\int_0^\infty \frac{{\rm d}q}{q}\left\{1\,-\,
2\int_q^\infty {\rm d}q_0\, q_0\,\delta\Bigl(q_0^2
-q^2-\Pi_t(q_0,q)\Bigr)\right\}.\eeq
In going to the second line, we have used the familiar sum-rule 
\cite{MLB96}\beq
\int_{-\infty}^\infty\frac{{\rm d}q_0}{2\pi}\, q_0\,
\rho_t(q_0,q)&=&1,\eeq
together with the parity property $\rho_t(-q_0,q)=-
\rho_t(q_0,q)$, to write
\beq
\int_{-q}^q \frac{{\rm d}q_0}{2\pi}\, q_0\, \beta_t(q_0,q)\,=\,1\,-
2\int_{q}^\infty \frac{{\rm d}q_0}{2\pi}\, q_0\,\rho_t(q_0,q)\,;\eeq
 then we have related the on-shell magnetic spectral
density to the plasmon pole in the transverse photon propagator:
$\rho_t(q_0>q)=2\pi \delta(q_0^2-q^2-\Pi_t(q_0,q))$.
We also recall that, in the hard thermal loop approximation,
 $\Pi_t(q_0,q)$ is a function of $q_0/q$ alone.

The integral over $q$ in eq.~(\ref{I20}) is well defined as it stands.
However, in order to work out separately the two terms within the braces,
 it is necessary to introduce, at intermediate steps,
an ultraviolet cutoff $\Lambda$ and also an infrared cutoff $\mu$.
The first term reads then
\beq\label{I21}
I_{21}(\Lambda,\mu)\,\equiv\,-\int_\mu^\Lambda \frac{{\rm d}q}{q}\,=\,
-\,\ln\frac{\Lambda}{\mu}\,.\eeq
The second term,
\beq\label{I22}
I_{22}(\Lambda,\mu)\,\equiv\,2\int_\mu^\Lambda \frac{{\rm d}q}{q}
\int_q^\infty {\rm d}q_0\, q_0\,\delta\Bigl(q_0^2-q^2-\Pi_t(q_0/q)
\Bigr),\eeq
involves an integral along the transverse plasmon dispersion
relation, $q_0=\omega_t(q)$ with 
$\omega_t^2(q)=q^2+\Pi_t(\omega_t/q)$.
To perform the integral, we use the following change of variables:
\beq\label{COV1}
x\equiv q_0/q,\qquad\,\,y\equiv q_0^2-q^2,
\qquad\,\,{\rm d}q\,{\rm d}q_0\,=\,\frac{{\rm d}x\,
 {\rm d}y}{2(x^2-1)}\,,\eeq
and get
\beq\label{I221}
I_{22}(\Lambda,\mu)=\int {\rm d}x \int {\rm d}y
\,\frac{x}{x^2-1}\,\delta(y-\Pi_t(x))=
\int_{x_m}^{x_M}{\rm d}x\, \frac{x}{x^2-1}\,=\,\frac{1}{2}\,
\ln\frac{x_M^2-1}{x_m^2-1}\,.\,\,\eeq
The integration limits $x_m$ and $x_M$ are obtained as follows:
As $q\to \mu$, $x\to \omega_t(\mu)/\mu$. The dispersion relation
$\omega_t(q)$ can be found, e.g., in Refs. \cite{Pisarski89a,BIO96,MLB96}.
For $\mu \to 0$,  $\omega_t(\mu)\to \omega_p$,
and $x\to x_M(\mu)=\omega_p/\mu$.
As $q\to \Lambda$ (with large $\Lambda \gg \omega_p$),
 $\omega_t^2(\Lambda)\simeq \Lambda^2
+3\omega_p^2/2$, and  $x\to x_m(\Lambda)=1 +3\omega_p^2/4\Lambda^2$.
Together with eqs.~(\ref{I21}) and (\ref{I221}), this gives
\beq\label{I2F}
I_2\,=\,I_{21}+I_{22}\,=\,-\,\ln\frac{\Lambda}{\mu}\,+\,\frac{1}{2}\,
\ln\frac{2\Lambda^2}{3\mu^2}\,=\,\frac{1}{2}\,\ln\frac{2}{3}\,.\eeq

The remaining integrals $I_1$ and $I_3$ are evaluated similarly.
In the process, we need the following sum rules \cite{Pisarski93,prd}
\beq\label{SumR}
\int_{-q}^q\frac{{\rm d}q_0}{2\pi q_0}\,
\beta_l(q_0,q)&=&\frac{1}{q^2}\,-\,\frac{1}{q^2+3\omega_p^2}\,-\,
2\int_q^\infty \frac{{\rm d}q_0}{q_0}\,\delta\Bigl(q^2+\Pi_l(q_0/q)
\Bigr),\nonumber\\
\int_{-q}^q\frac{{\rm d}q_0}{2\pi q_0}\,
\nu_t(q_0,q)&=&\frac{1}{q^2+\omega_p^2}\,-\,
2\int_q^\infty \frac{{\rm d}q_0}{q_0}\,\delta\Bigl(q_0^2-q^2-\Pi_t(q_0/q)
\Bigr),\eeq
where $\Pi_l$ and $\Pi_t$ are the polarisation
functions in the hard thermal loop approximation.
(We use the same notations as in Ref. \cite{prd}.)
In the computation of $I_3$ --- which involves $\nu_t(q_0,q)$ ---
we change the integration variables as in eq.~(\ref{COV1}) 
above, and obtain:
\beq
I_3\,=\,\frac{1}{2}\,\ln\frac{3}{2}\,,\eeq
which happens to cancel $I_2$, eq.~(\ref{I2F}). 
As for the electric piece $I_1$, we write
\beq\label{I10}
I_1&=&- \int_0^\infty {\rm d}q \,q\left\{
\frac{1}{q^2}\,-\,\frac{1}{q^2+3\omega_p^2}\,-\,
2\int_q^\infty \frac{{\rm d}q_0}{q_0}\,\delta\Bigl(q^2+\Pi_l(q_0/q)
\Bigr)\right\}
\nonumber\\
&=&\ln\frac{\sqrt{3} \omega_p}{\mu}\,-\,2\int_\mu^\infty {\rm d}q \,q
\int_q^\infty \frac{{\rm d}q_0}{q_0}\,\delta\Bigl(q^2+\Pi_l(q_0/q)
\Bigr),\eeq
where an infrared cutoff $\mu$ was introduced when separating the terms
inside the braces. In the second term, we change the variables according
to \beq\label{COV2}
x\equiv q_0/q,\qquad\,\,y\equiv q^2,
\qquad\,\,{\rm d}q\,{\rm d}q_0\,=\,\frac{1}{2}
\,{\rm d}x\, {\rm d}y\,,\eeq
and get
\beq\label{I12}
-\int_1^{x_M}\frac{{\rm d}x}{x} \int {\rm d}y
\,\delta(y+\Pi_l(x))\,=\,-
\ln x_M\,=\,-\ln \frac{\omega_p}{\mu}\,.\eeq
The upper limit was obtained as $x_M(\mu)=\omega_l(\mu)/\mu
\simeq \omega_p/\mu$ for $\mu \to 0$.
From eqs.~(\ref{I10}) and (\ref{I12}), we finally obtain
\beq
I_1\,=\,\frac{1}{2}\,\ln 3,\eeq
which is the result quoted in eq.~(\ref{FregF}).

We finally evaluate the momentum integral
in eq.~(\ref{CV}) in the non-relativistic limit $v\ll 1$.
Since  $|q_0|\le vq\ll q$, we need
the spectral functions $\beta_{l,\,t}(q_0,q)$
only for very small frequencies \cite{Pisarski89a} :
\beq\label{RHOS}
\beta_{l}(q_0\ll q)&\simeq&\frac{3\pi\omega_p^2\,(q_0/q)}
{(q^2 + 3\omega_p^2)^2}\nonumber\\
\beta_{t}(q_0 \ll q)&\simeq&\frac{3\pi\omega_p^2\,(q_0/2q)}
{q^4 + (3\pi \omega_p^2 q_0/4q)^2}\,.\eeq
 Corresponding to the three terms in 
 eq.~(\ref{CV}), we write $C(v)=C_1(v)+C_2(v)+C_3(v)$.

The electric contribution is evaluated as follows:
\beq\label{C1}
C_1(v)\,=\,\frac{1}{v^2}\int_0^\infty {\rm d}q \,q
 \int_{-vq}^{vq} \frac{{\rm d}q_0}{2\pi q_0}
\beta_l(q_0,q)\,\simeq\,\frac{1}{v^2}\int_0^\infty {\rm d}q \,q
\frac{3\omega_p^2\,v}{(q^2 + 3\omega_p^2)^2}\,=\,
\frac{1}{2v}\,,\eeq
where the neglected
terms are smaller, at least, by two powers of $v$
(since $\beta_{l}(q_0,q)$ is an odd function of $q_0$).

The first magnetic contribution is
\beq
C_2(v)&=&-\,\frac{1}{v^2}\int_0^\infty {\rm d}q \,q
 \int_{-vq}^{vq} \frac{{\rm d}q_0}{2\pi q_0}
\,\frac{q_0^2}{q^2}\,\beta_t(q_0,q)\nonumber\\
&\simeq&-\,\frac{8}{3\pi^2  \omega_p^2 v}
\int_0^\infty {\rm d}q \,q\left[1\,-\,\frac{\arctan y(q;v)}
{ y(q;v)}\right],\eeq
where we have used the approximate expression (\ref{RHOS})
for  $\beta_t(q_0,q)$ to perform the integral over $q_0$,
and we have denoted $ y(q;v)\equiv 3\pi\omega_p^2\,v/(4q^2)$.
In the remaining integral over $q$, we make the obvious change
of variables $ y(q;v) \equiv t$, with ${\rm d}q/q\,=\,
{\rm d}t/(2t)$, and obtain
\beq\label{C2}
C_2(v)\,=\,\frac{1}{\pi}\int_0^\infty\frac{{\rm d}t}{t^2}
\left[1\,-\,\frac{\arctan t}{t}\right]
\,=\,\frac{1}{\pi}\int_0^\infty {\rm d}y \,[1\,-\,y\,{\rm arccot} y]
\,=\,\frac{1}{4}\,,\eeq
which is independent of $v$.

Finally, the second  magnetic contribution reads
\beq\label{C31}
C_3(v)\,=\,\int_0^\infty {\rm d}q \,q
 \int_{-vq}^{vq} \frac{{\rm d}q_0}{2\pi q_0}\,
\nu_t(q_0,q)\, 
\simeq\,\int_0^\infty\frac{{\rm d}q}{q}\left[\frac{2}{\pi}
\arctan y(q;v)\,-\,\frac{\omega_p^2}{q^2+\omega_p^2}
\right],\eeq
where we have also used the definition (\ref{sept})
of $\nu_t(q_0,q)$. Note that, for any $v>0$, the remaining
integral over $q$ is well-defined, and saturated
by soft momenta, $q\simle \omega_p$. Still, the
limit $v\to 0$ is not well-defined (because of potential
infrared singularities), so that we need to perform
the momentum integral before studying the small $v$
behaviour. By using the same change of variables
as above, we rewrite eq.~(\ref{C31}) as
\beq\label{C32}
C_3(v)&\simeq&\int_0^\infty\frac{ {\rm d}t}{2t}
\left[\frac{2}{\pi}\,\arctan t\,-\,\frac{t}
{t +\tilde v}\right]\nonumber\\
&=&\int_0^1{\rm d}t\,\frac
{\arctan t}{\pi t}\,+\,\int_1^\infty\frac{ {\rm d}t}{\pi t}
\left[\arctan t\,-\,\frac{\pi}{2}\right]+\,\frac{1}{2}\,
\ln\tilde v\,,\eeq
with $\tilde v \equiv 3\pi v/4$. 
The two integrals in the second line mutually cancel,
as can be seen by changing $t\to 1/t$ in any of them,
and then  using $\arctan 1/t = \pi/2 - \arctan t$. Finally,
\beq\label{C3}
C_3(v)&\simeq&\frac{1}{2}\ln\tilde v\,.\eeq
By putting together the above results in eqs.~(\ref{C1}),
(\ref{C2}) and (\ref{C3}), one obtains the result
quoted in eq.~(\ref{CVNR}).

\setcounter{equation}{0}
\vspace*{2cm}
\renewcommand{\theequation}{C.\arabic{equation}}
\appendix{\noindent {\large{\bf Appendix C}}}

In section 5 above, we have used a non-Abelian version of the
Bloch-Nordsieck model to study the interactions between
hard quasiparticles (quarks or gluons) and soft virtual gluons
in hot QCD. In this appendix, we examine the validity
of this approximation  for the case where the hard quasiparticle
is a transverse gluon.

\begin{figure}
\protect \epsfxsize=15.6cm{\centerline{\epsfbox{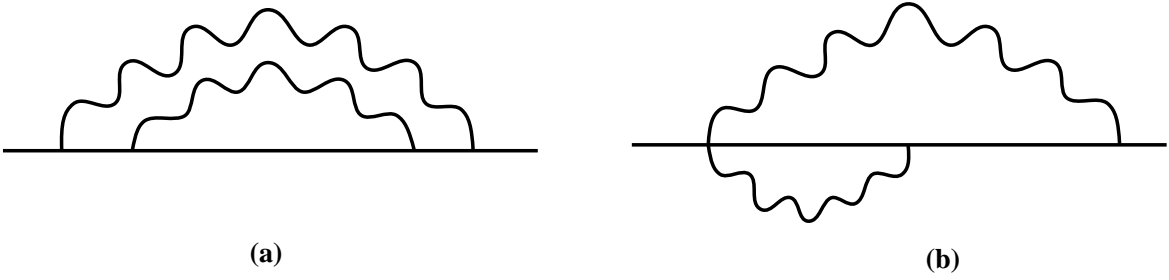}}}
	 \caption{Some two-loop self-energy corrections in QCD.
The continuous line represents a hard gluon, which is nearly
on-shell. The wavy lines are soft magnetostatic gluons.
In the on-shell limit, diagram (b), which also involves
 a four-gluon vertex, is less infrared singular then diagram (a).}
\label{2QCD}
\end{figure}

\begin{figure}
\protect \epsfxsize=14.cm{\centerline{\epsfbox{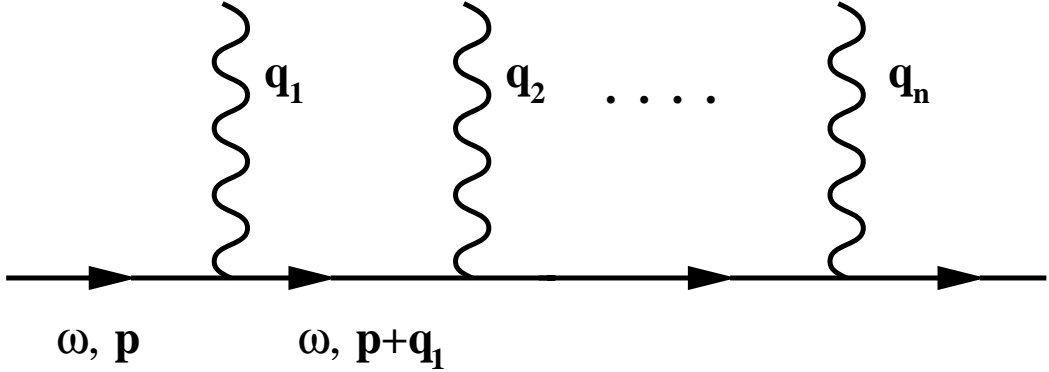}}}
	 \caption{A typical diagram contributing to $G(x,y|{\bf A})$
to order $g^n$ in perturbation theory. This diagram involve $n$ gluon
field insertions, and $n+1$ free propagators $G_0$ (including
the external lines). The external fields are purely static and magnetic.}
\label{BNfig}
\end{figure}

The dominant contributions to the hard ($p\simge T$) gluon propagator
near the mass-shell at $\omega = p$
(i.e., the leading infrared divergences for $\omega \to p$)
come from the diagrams illustrated in Fig.~\ref{QCD}.
The continous line there represents the hard gluon
and the wavy lines denote very soft ($q\ll gT$) static ($q_0=0$)
magnetic gluons. Once again, our strategy is to consider
first the interactions with a classical, static, color field 
${\bf A}^a({\bf q})$. Then, a typical diagram looks like
in Fig.~\ref{BNfig}. The self-energy corrections in Fig.~\ref{QCD}
will be eventually recovered by functional integration, as shown
in eq.~(\ref{FS}).

A noteworthy feature of Figs.~\ref{BNfig} and \ref{QCD} is that
the hard particle is involved only in three-gluon (but not
in four-gluon) vertices. (Of course, the four-gluon vertices do
also enter the self-energy diagrams --- see, e.g., Fig.~\ref{QCD} ---, 
but they couple only soft internal gluons; cf. eq.~(\ref{FS}).)
The reason is that, to a given order in perturbation theory,
the diagrams which involve the hard particle in 
four-gluon vertices are less infrared singular. This can be easily
verified by power counting: Consider, e.g., the two two-loop graphs
in Fig.~\ref{2QCD}. For $\omega=p$, the diagram in Fig.~\ref{2QCD}.a,
with only three-gluon vertices, gives rise
to a linear infrared singularity. That is, its contribution
to the damping rate is of the order $\gamma^{(2a)}\sim g^4 T^2/\mu$ 
(up to logarithms of $gT/\mu$), which for $\mu\sim g^2T$
gives $\gamma^{(2a)}\sim g^2T$; i.e.,
it is of the same order as the one-loop contribution.
(This leading divergence can be isolated by using the 
simplified BN Feynman rules to be derived below.
See Appendix C in Ref. \cite{prd} for a detailed analysis.)
The diagram in Fig.~\ref{2QCD}.b, which also involves one four-gluon vertex,
may give rise, at most, to logarithmic mass-shell singularities. 
We thus expect $\gamma^{(2b)}\sim g^4 T^2/p \sim g^4 T$,
which stands beyond our present accuracy, and should be 
discarded for consistency.
We shall verify shortly that, for the problem at hand, neglecting
 the four-gluon vertices is indeed consistent with gauge symmetry.

Consider the diagram in  Fig.~\ref{BNfig}, with only three-gluon
vertices. The latter are linear in the external gluon momenta:
 \beq\label{TL3G}
-i gf^{abc}\,\Gamma_{ijl}({\bf p, q, k})=-igf^{abc}\Bigl(
(p-q)_l \delta_{ij} + (q-k)_i \delta_{jl} +
(k-p)_j \delta_{il}\Bigr),\eeq 
where ${\bf p+q+k}=0$. 
Remember that all the external lines  in  Fig.~\ref{BNfig}
are of the magnetic type, so that we need just
the spacial components of the vertex function.
Furthermore, color indices play no special role for the subsequent 
kinematic approximations, and will be omitted in intermediate
formulae.

For all the vertices in  Fig.~\ref{BNfig}, one of the external momenta 
is soft, since it is carried by the classical color field.
If ${\bf q}$ is the soft momentum in eq.~(\ref{TL3G}), then
\beq\label{Gsim}
\Gamma_{ijl}({\bf p, q, k})\,\simeq\,\Gamma_{ijl}({\bf p}, 0,{\bf -p})\,
=\,p_l \delta_{ij} + p_i \delta_{jl} - 2 p_j \delta_{il}.\eeq 
Since the approximate three-gluon vertex (\ref{Gsim})
is independent of the soft momentum ${\bf q}$, the Ward identities
are consistent with setting the four-gluon vertex to zero,
which is what we did before.

Consider now a typical internal gluon line
in  Fig.~\ref{BNfig}: It is necessarily hard
and nearly on-shell. In the Coulomb gauge, the
 associated propagator reads (recall that $q_0=0$)
\beq\label{D00}
D_{ij}(\omega, {\bf p+q})\,=\,\frac {\delta_{ij} -{(\hat p_i+
\hat q_i)(\hat p_j+\hat q_j)}}
{\omega^2-({\bf p+q})^2}\,
\simeq\,
\frac{1}{2p}\,\frac {\delta_{ij} -{v_i v_j}}
{\omega  - {\bf v\cdot (p+ q)}}\,,\eeq
where $\hat p_i=p_i/p\equiv v_i$  and
the approximate equality holds since $q\ll p$ and $\omega \sim p$.
That is, $D_{ij}(\omega, {\bf p+q})\simeq (1/2p) {\cal P}_{ij}\,
G_0(\omega, {\bf p+q})$, where $ {\cal P}_{ij}=\delta_{ij} -v_i v_j$
is a transverse projector, and $G_0$ is the BN propagator
(cf. eq.~(\ref{G0})):
\beq\label{G01}
G_0(\omega, {\bf p+q})\,=
\,\frac{1}{\omega -{\bf v}\cdot {\bf (p+q)}}\,.\eeq
In Fig.~\ref{BNfig}, all the three-gluon vertices like (\ref{Gsim})
appear between projectors like $ {\cal P}_{ij}$.
By using the identity
\beq
 {\cal P}_{im}\,\Gamma_{mjn}({\bf p}, 0,{\bf -p})\, {\cal P}_{nl}
\,=\,-\,2(\delta_{il} -v_i v_l)p_j\,=\,-\,
2p\, {\cal P}_{il}\,v_j,\eeq
it can then be easily verified that the leading contribution
of the diagram \ref{BNfig} to the hard gluon propagator
can be evaluated with the following simplified Feynman rules
(we reintroduce here the color indices):
 (i) the hard particle propagator $\delta^{ab}
G_0(\omega, {\bf p+q})$, and (ii) the hard particle-soft gluon
vertex $igf^{abc} v_i$.
These are the Feynman rules which have been used 
to define the non-Abelian Bloch-Nordsieck model in section 6.
 For a hard quark, the 
color indices in the above Feyman rules should be replaced by
the corresponding indices in the fundamental representation.

\end{document}